\documentclass[a4paper,11pt]{article}
\pdfoutput=1 

\usepackage[fleqn]{amsmath}
\usepackage{algorithm,algorithmic,amssymb,amsthm,color,colortbl,cleveref,enumerate,graphicx,multirow,stmaryrd,url}

\newcommand{\COMPLEXITYCLASSNAMESTYLE}[1]{\mathsf{#1}}
\newcommand{\PROBLEMNAMESTYLE}[1]{\textsc{#1}}
\newcommand{\PROBLEMNAMEDELIMITER}{\hspace{1mm}}

\newcommand{\BOUNDED}{\PROBLEMNAMESTYLE{Bounded}}
\newcommand{\CLASSP}{\COMPLEXITYCLASSNAMESTYLE{P}}
\newcommand{\CONP}{\COMPLEXITYCLASSNAMESTYLE{co}\NP}
\newcommand{\COPT}{\PARTSTYLE{C1}}
\newcommand{\CTPT}{\PARTSTYLE{C2}}
\newcommand{\DELIM}{\#}
\newcommand{\DUMMY}{d}
\newcommand{\DOT}{\hspace{-0.1em}\cdot\hspace{-0.1em}}
\newcommand{\FP}{\mathsf{FP}}
\newcommand{\FROB}{\PROBLEMNAMESTYLE{Frobenius}}

\newcommand{\GAP}{\PROBLEMNAMESTYLE{\NUMERICALSEMIGROUP{\PROBLEMNAMEDELIMITER}{\PROBLEMNAMEDELIMITER}Gaps}}
\newcommand{\IGAP}{\PROBLEMNAMESTYLE{NonRep}}

\newcommand{\MOD}{\,\mathrm{mod}\,}
\newcommand{\NP}{\COMPLEXITYCLASSNAMESTYLE{NP}}

\newcommand{\NUM}{\PROBLEMNAMESTYLE{\#}}
\newcommand{\NUMD}{\PROBLEMNAMESTYLE{\#}\hspace{-0.4mm}\cdot\hspace{-0.2mm}}
\newcommand{\NUMERICALSEMIGROUP}[1]{NS}
\newcommand{\OPT}{\mathsf{Opt}}
\newcommand{\OSPT}{\PARTSTYLE{1S}}
\newcommand{\OT}{\PROBLEMNAMESTYLE{1IN3}}

\newcommand{\PARTSTYLE}[1]{\mathit{#1}}
\newcommand{\PH}{\COMPLEXITYCLASSNAMESTYLE{PH}}
\newcommand{\PI}[1]{{\COMPLEXITYCLASSNAMESTYLE{\Pi}_{#1}}}
\newcommand{\PIP}[1]{{\COMPLEXITYCLASSNAMESTYLE{\Pi}_{#1}^{\COMPLEXITYCLASSNAMESTYLE{P}}}}
\newcommand{\PSPACE}{\CLASSP\SPACE}
\newcommand{\QED}{\hfill$\square$}
\newcommand{\SAT}{\PROBLEMNAMESTYLE{SAT}}
\newcommand{\SIGMA}[1]{{\COMPLEXITYCLASSNAMESTYLE{\Sigma}_{#1}}}
\newcommand{\SIGMAP}[1]{{\COMPLEXITYCLASSNAMESTYLE{\Sigma}_{#1}^{\COMPLEXITYCLASSNAMESTYLE{P}}}}
\newcommand{\SPACE}{\COMPLEXITYCLASSNAMESTYLE{SPACE}}

\newcommand{\TGAP}{\BOUNDED\PROBLEMNAMEDELIMITER\GAP}

\newcommand{\VAR}{\mathrm{Var}}
\newcommand{\XAPT}{\PARTSTYLE{XA}}
\newcommand{\XOPT}{\PARTSTYLE{X1}}
\newcommand{\XTPT}{\PARTSTYLE{X2}}
\newcommand{\YOPT}{\PARTSTYLE{Y1}}
\newcommand{\YTPT}{\PARTSTYLE{Y2}}

\newtheorem{definition}{Definition}
\newtheorem{theorem}[definition]{Theorem}
\newtheorem{claim}[definition]{Claim}
\newtheorem{corollary}[definition]{Corollary}
\newtheorem{example}[definition]{Example}

\newtheorem{problem}[definition]{Problem}
\newtheorem{remark}[definition]{Remark}

\crefname{claim}{Claim}{Claims}
\crefname{corollary}{Corollary}{Corollaries}
\crefname{example}{Example}{Examples}
\crefname{lemma}{Lemma}{Lemmata}
\crefname{problem}{Problem}{Problems}
\crefname{remark}{Remark}{Remarks}
\crefname{subsection}{subsection}{Subsections}
\crefname{subsubsection}{subsubsection}{Subsubsections}
\crefname{theorem}{Theorem}{Theorems}

\newcommand{\OURTITLE}{The Complexity of the Numerical Semigroup Gap Counting Problem}
\newcommand{\OURNAME}{Shunichi Matsubara}

\definecolor{MyGray}{cmyk}{0.00,0.00,0.00,0.50}
\definecolor{MyGrayLight}{cmyk}{0.00,0.00,0.00,0.18}

\begin{document}
\title{\OURTITLE}
\author{\OURNAME\\
Aoyama Gakuin University\\
Japan
\thanks{matsubara@it.aoyama.ac.jp}}
\date{}

\maketitle

\begin{abstract}
In this paper, we prove that the numerical-semigroup-gap counting problem 
is $\NUM\NP$-complete as a main theorem.
A numerical semigroup is an additive semigroup over the set of all nonnegative integers. 
A gap of a numerical semigroup is defined as a positive integer that does not belong to the numerical semigroup.
The computation of gaps of numerical semigroups has been actively studied from the 19th century.
However, little has been known on the computational complexity.
In 2005, 
Ram\'irez-Alfons\'in proposed a question whether or not
the numerical-semigroup-gap counting problem is $\NUM\CLASSP$-complete.
This work is an answer for his question.
For proving the main theorem, 
we show the $\NUM\NP$-completenesses of
other two variants of the numerical-semigroup-gap counting problem.
\end{abstract}

\section{Introduction}
\label{sec:intro}

The numerical-semigroup-gap counting problem%
~\cite{ramirezAlfonsin2005diophantine,rosales2009numerical,Sylvester1882AmJMath5}
has been actively researched since 19th century
from the mathematical point of view.
We denote this problem by $\NUM\GAP$.
This problem is defined as follows.
Given a set $A$ of coprime positive integers $a_1,\cdots,a_n$,
the task is counting the number of all integers that cannot be represented
as nonnegative integer combinations of $A$.
The set of those integers
consists of an additive semigroup called a {\itshape numerical semigroup}~\cite{rosales2009numerical} 
and denoted $\mathcal{S}(A)$.
An integer not in $\mathcal{S}(A)$ is called a {\itshape gap} of $\mathcal{S}(A)$
and the set of all gaps of $\mathcal{S}(A)$ is denoted $N(A)$.
An integer in $\mathcal{S}(A)$ is called a {\itshape nongap} of $\mathcal{S}(A)$.
For example, if $A=\{6,11,15\}$ are given,
then $N(A)$ consists of $10$ integers $1, 2, 3, 4, 7, 8, 9, 13, 14, 19$.
For a given set $A$ of coprime positive integers,
it is known that $N(A)$ is finite~\cite{rosales2009numerical,ramirezAlfonsin2005diophantine}.
The maximum integer of $N(A)$ is called the {\itshape Frobenius number} and denoted $g(A)$.

\subsection{Computation of the numerical-semigroup-gap counting problem}
\label{subsec:1_variants}

In this paper, we assume that the reader has some acquaintance
with the fundamentals of computational complexity theory.
The reader is referred to textbooks; e.g.,
\cite{arora2009computational,du2011theory,papadimitriou1995computational}.
We denote the class of all polynomial-time computable function (search) problems by $\FP$.
In this paper, given an input,
we assume its bit length to be the parameter.
For example, given a finite set $A$ of positive integers as an input of a problem,
the number $\sum_{a \in A}(\lfloor \log a \rfloor + 1)$ is its input size.

Little research has been known 
from the complexity theoretical point of view for the problem $\NUM\GAP$
while many results have been known from others.
Ram\'irez-Alfons\'in proposed the following question
(\cite{ramirezAlfonsin2005diophantine}, Problem A.4.2).
{\itshape ``Is computing $N(a_1,\cdots,a_n)$ $\NUM\mathcal{P}$-complete?''}
(In \cite{ramirezAlfonsin2005diophantine},
Ram\'irez-Alfons\'in uses the notation $N(A)$
for describing two different concepts;
the set of all gaps of $\mathcal{S}(A)$ and its cardinality,
where $A$ is a finite set of coprime positive integers.
From the context, it is clear to denote the latter in the above question.)
This problem has remained open until the present work.
This paper gives an answer, which is negative unless $\NUM\CLASSP = \NUM\NP$.
$\NUM\CLASSP$ is one of the best-known classes of counting problems~\cite{VALIANT1979TCS}
and the counting counterpart of $\NP$.
We formally review the class $\NUM\CLASSP$ formally in Section~\ref{subsec:2_complexity_class}.

On computation of $\NUM\GAP$, more results are known.
Sylvester \cite{Sylvester1882AmJMath5} showed $|N(a_1,a_2)|$ to be $(a_1-1)(a_2-1)/2$
for two coprime positive integers $a_1$ and $a_2$.
(\cite{ramirezAlfonsin2005diophantine}, Section 5)
surveys results on equalities and inequalities for $\NUM\GAP$.
Barvinok~\cite{barvinok2003short} found a polynomial-time algorithm
for counting the number of elements in $N(A)$
under the assumption that the cardinality of $A$ is fixed.
In more detail, 
Barvinok~\cite{barvinok2003short} showed a short rational function
for a generating function that counts all the elements of $N(A)$.
Unfortunately, 
this algorithm is doubly exponential in the cardinality $n$ of $A$.

\subsection{Variants of the numerical-semigroup-gap counting problem}
\label{subsec:intro_2}

This paper concerns counting all elements of $N(A)$ for a given 
set $A$ of coprime positive integers.
This problem is closely related to the problem finding the maximum element of $N(A)$,
the Frobenius number $g(A)$, 
for a given set $A$ of coprime positive integers.
This problem has been studied as actively as on $\NUM\GAP$.
In this paper, 
we call this decision problem the {\itshape Frobenius problem}
and denote it by $\FROB$
although in general, the problem is often called by several different names;
the linear Diophantine problem of Frobenius,
the coin exchange problem of Frobenius, 
the money changing problem, 
and so on.
Ram\'irez-Alfons\'in~\cite{ramirezAlfonsin1996ComplexityOfFP} 
proved $\FROB$ to be $\NP$-hard under Cook reductions.
More recently, Matsubara~\cite{matsubara2016DecisionFrobeniusComplexity}
improved this lower bound.
In more detail, he proved the $\SIGMAP{2}$-hardness of the Frobenius problem
under Karp reductions.
We formally review the class $\SIGMAP{2}$ in Section~\ref{subsec:2_complexity_class}.
These results suggests that the problem $\NUM\GAP$ is at least as has as
$\NUM\CLASSP$.

However, if we assume the number of input integers to be fixed,
then $\FROB$ is known to be polynomial-time computable.
Kannan~\cite{Kannan1992} found a polynomial-time algorithm
for solving the Frobenius problem under the assumption that 
the number of input integers is fixed.
Barvinok~\cite{barvinok2003short}
also found a polynomial-time algorithm under the same assumption.

Many experimentally fast algorithms were developed
for solving $\FROB$; e.g.,
\cite{beihoffer2005faster,einstein2007frobenius,Roune20081}, 
Section 1 in \cite{ramirezAlfonsin2005diophantine}.

\subsection{Complexity classes and polynomial-time reductions}
\label{subsec:intro_3}

In this paper, we concern computational complexity of counting.
In particular, we prove $\NUM\GAP$ and its variants to be complete 
for $\NUM\NP$ under two types of polynomial-time reductions,
called parsimonious and relaxed subtractive reductions.
The class $\NUM\NP$ can be considered to be the counting counterpart of $\SIGMAP{2}$.
A relaxed subtractive reduction is a new type of reduction introduced in this paper.

Meyer and Stockmeyer~\cite{MeyerStockmeyer1972}
found a well-known hierarchy of decision problems, called
the polynomial-time hierarchy and
many results are known on
this hierarchy~\cite{schaefer2002completeness,schaefer2002completenessII,Stockmeyer19761,Wrathall197623}.
This hierarchy is contained in $\PSPACE$. 
We review this hierarchy formally in Section~\ref{sec:preliminary}.

Valiant\cite{VALIANT1979TCS,Valiant1979SICOMP} introduced $\NUM\mathsf{P}$ 
and developed the theory of the complexity of counting.
Many complete problems are found for $\NUM\CLASSP$~\cite{Valiant1979SICOMP}.
In any problem $\NUM L \in \NUM\CLASSP$,  
we count the number of witnesses for a given instance
while in any problem $L \in \NP$,  
we check the existence of a witness for a given instance.
The relationship between counting complexity classes
$\FP$ and $\NUM\CLASSP$ does not entirely correspond to 
the one of the decision complexity classes
$\CLASSP$ and $\NP$, respectively.
In more detail, 
there is a decision problem $L \in \CLASSP$
such that its counting counterpart is $\NUM\CLASSP$-complete.
For example, the permanent problem~\cite{VALIANT1979TCS} is $\NUM\CLASSP$-complete
while its decision counterpart is the bipartite perfect matching problem, which is in $\CLASSP$.
This phenomenon is often called {\itshape ``easy to decide but hard to count''}.

The descriptive power of $\NUM\CLASSP$ is known to be so strong.
Toda proved the polynomial hierarchy to be Cook reducible
to $\NUM\CLASSP$~\cite{Toda1991SICOMP}.
However, some natural counting problems appear to 
be not in $\NUM\CLASSP$~\cite{Durand2005496,Hermann2010634,Hermann20093814}.
Under such a background,
several types of counting complexity classes were introduced.
In this paper, we use classes in the following two types of counting classes
$\NUM\SIGMAP{k}$~\cite{VALIANT1979TCS} 
and $\NUMD{\SIGMAP{k}}$~\cite{TodaPhD1991English},
which are classes in counting variants of the polynomial hierarchy. 
As the complement of the latter hierarchy, we also use $\NUMD{\PIP{k}}$.
We formally define these classes in Section~\ref{subsec:2_complexity_class}.

A counting problem $\NUMD{\SIGMA{k}\SAT}$
is a representative of counting problems in a class wider than $\NUM\CLASSP$, 
which is a counting variant of the $(k+1)$-alternating quantified Boolean satisfiability problem.
Moreover, $\NUMD{\SIGMA{k}\SAT}$ is shown to be $\NUMD{\SIGMAP{k}}$-complete~\cite{Durand2005496}.
Reductions of problems in $\NUMD{\SIGMA{k}\SAT}$ and $\NUMD{\PI{k}\SAT}$
are sensitive for which type of reduction we adopt.
Parsimonious reductions are weaker reductions for counting problems.
$\NUMD{\SIGMA{k}\SAT}$ and $\NUMD{\PI{k}\SAT}$ are closed under
parsimonious reductions. 
If two counting problems have similar structures, 
then we can find a parsimonious reduction. 
Unfortunately, it is not easy to find a parsimonious reduction for given problems
in many cases.
We review parsimonious reductions in Section~\ref{subsec:2_reducibilities}.
On the other hand, Cook reductions are too strong to reduce 
counting problems in the following sense.
Toda and Watanabe~\cite{TODA1992205} showed that the hierarchy
$\NUMD{\SIGMAP{k}}$ is not closed under Cook reductions 
even if we restrict the number of its oracle calls to at most once.
Durand et al.~\cite{Durand2005496} introduced subtractive reductions. 
Subtractive reductions are intermediates between parsimonious and Cook reductions.
They found some complete counting problems
for $\NUMD{\PIP{k}}$ under subtractive reductions.

Schaefer~\cite{Schaefer:1978:CSP:800133.804350} found a dichotomy theorem
on the general satisfiability problems.
By this theorem,
we can classify
all general satisfiability problems into one of $\CLASSP$ and $\NP$-complete problems 
by some properties.
Creignou and Hermann~\cite{Creignou19961} found the counting version 
of the dichotomy theorem for the general satisfiability problems.
By this theorem, 
we can classify all the general satisfiability counting problems
into one of $\FP$ and $\NUM\CLASSP$-complete problems.
Moreover, 
Bauland et al.~\cite{BaulandBohlerCreignouReithSchnoorVollmer2010TOCS}
found a trichotomy theorem for the class of counting problems. 
This theorem is a generalization of the dichotomy theorem in~\cite{Creignou19961}
to the classes $\NUMD{\SIGMAP{k}}$ and $\NUMD{\PIP{k}}$.
They introduced another type of reduction, called a complementive reduction,
and showed the trichotomy that classifies any $k$-alternating quantified satisfiability counting problem
into a $\FP$ problem, a $\NUM\CLASSP$-complete problem under Cook reductions, 
and a $\NUMD{\SIGMAP{k}}$-complete problem under complementive reductions for some $k \geq 1$. 
Note that, as described by \cite{BaulandBohlerCreignouReithSchnoorVollmer2010TOCS}, 
the {\itshape ``$\NUMD{\SIGMAP{k}}$-completeness under complementive reductions''}
does not have a more desirable property in structural complexity theory.
In more detail, $\NUMD{\SIGMAP{k}}$ is not closed under complementive reductions,
although $\NUMD{\PIP{k}}$ is closed under complementive reductions.

Hermann and Pichler~\cite{Hermann2010634} proved some counting problems
on propositional abduction to be complete
for some classes that are at least as hard as $\NUM\CLASSP$.
In that work,
some of the completenesses are proved under subtractive reductions.
Hermann and Pichler~\cite{Hermann20093814}
also investigated some problems that count optimal solutions for given instances
by using the counting complexity class $\NUMD{\OPT_k\CLASSP}$.
The class $\NUMD{\OPT_k\CLASSP}$
is an extension of the class introduced by Krentel~\cite{KRENTEL1992183}.
Hemaspaandra and Vollmer~\cite{Hemaspaandra:1995:SNC:203610.203611}
surveyed results on several counting classes wider than $\NUM\CLASSP$.

\subsection{Contributions of this work}

The main contribution of this work
is the proof for the $\NUM\NP$-completeness of $\NUM\GAP$.
This result is also an answer for the open problem
proposed by Ram\'irez-Alfons\'in, 
which is a negative answer unless $\NUM\NP = \NUM\CLASSP$.
That completeness will be proved under 
a new type of reduction, which we call a relaxed subtractive reduction.
For the proof of the main theorem,
we will prove the $\NUM\NP$-completenesses of two variants of $\NUM\GAP$.
One is a problem $\NUM\IGAP$. 
The task of $\NUM\IGAP$ is counting the number of all positive integers 
that are in a given interval and cannot be represented
as nonnegative integer combinations of a given set of positive integers.
The other is $\NUM\TGAP$.
The task of $\NUM\TGAP$ is 
counting the number of all positive integers 
that are greater than or equal to a given bound 
and cannot be represented as nonnegative integer combinations of a given set of positive integers.
We will show the $\NUM\NP$-completenesses of
those two variants under parsimonious reductions.
Our proofs are self-contained for the readability.
We prove almost all statements 
by using combinatorial methods
and elementary properties on numerical semigroup.
We do not use the existing notions and results in numerical semigroup
if not necessary.

\subsection{Organization of the paper}

In Section~\ref{sec:preliminary},
we define notions and notations
for the proofs of the later sections.
In Sections~\ref{subsec:2_2} and \ref{subsec:2_complexity_class},
we define the basic notions and notations on
Boolean satisfiability and computational complexity, respectively.
In Section~\ref{subsec:2_reducibilities},
we introduce relaxed subtractive reductions. 
In Section~\ref{subsec:computational_problems},
we list related computational problems.
Section~\ref{subsec:2_sat_completeness} describes the $\NUM\NP$-completeness of 
$\NUMD{\PI{k}\OT\SAT}$, a counting problem on Boolean satisfiability,
under parsimonious reductions.
Then, the remaining sections,
we prove the $\NUM\NP$-completenesses of 
three variants of the numerical-semigroup-gap counting problem.
In Section~\ref{sec:proof_parsimonious_reducibility},
we prove the $\NUM\NP$-completeness of $\NUM\IGAP$
under parsimonious reductions.
In Section~\ref{sec:proof_bounded_gaps_parsimonious_reducibility},
we prove the $\NUM\NP$-completeness of $\NUM\TGAP$
under parsimonious reductions.
In Section~\ref{sec:proof_relax_subtractive_reducibility},
we prove the $\NUM\NP$-completeness of $\NUM\GAP$
under relaxed subtractive reductions.
Section~\ref{sec:future_work_and_concluding_remarks}
concludes this paper with future work.

\section{Preliminaries}
\label{sec:preliminary}

We denote by $\mathbb{N}$ and $\mathbb{N}_+$
the sets of all nonnegative integers and all positive integers, respectively.
For any two nonnegative integers $a$ and $b$ with $a \leq b$,
we denote the interval $\{c \in \mathbb{N} \colon a \leq c \leq b\}$
by $[a,b]$.
For any integer $k$, 
we denote the set $\{c \in \mathbb{N} \colon c \geq k\}$
by $[k,\infty]$.
In this paper,
for any integers $a_1,\cdots,a_n$,
we consider a nonnegative integer combination
$\sum_{i=1}^n c_i a_i$
to be the sum of integers $a_{i_1},\cdots,a_{i_l}$,
where $c_i \in \mathbb{N}$ for every $1 \leq i \leq n$
and $1 \leq i_1 \leq \cdots \leq i_l \leq n$.
For any $n \in \mathbb{N}_+$,
we denote the number of $1$s in the binary representation of $n$
by $\#_1(n)$.

Let $A$ denote a set of positive integers $a_1,\cdots,a_n$.
In Section~\ref{sec:intro},
when $a_1,\cdots,a_n$ are coprime, 
we defined $\mathcal{S}(A)$ as the numerical semigroup,
and defined $N(A)$ as the set of all gaps of $\mathcal{S}(A)$.
We extend these notations to any set of positive integers as follow.
For any $A$, we denote the set
$\{\sum_{a \in A} c_a a \colon c_a \in \mathbb{N} \}$
by $\mathcal{S}(A)$,
and the set $\mathbb{N} \backslash \mathcal{S}(A)$
by $N(A)$.
Needless to say, if $a_1,\cdots,a_n$ are not coprime integers,
then $\mathcal{S}(A)$ is not a numerical semigroup.

We identify any integer as its binary representation
if no confusion arises.
We denote a characteristic function on a predicate $p$
by $\llbracket p \rrbracket$; i.e.,
$\llbracket p \rrbracket$ is $1$ if $p = 1$,
and $0$ otherwise.
The notation $\llbracket \cdot \rrbracket$ 
was introduced by Kenneth E. Iverson,
and often called {\itshape ``Iverson bracket''}~(\cite{Graham:1994:CMF:562056}, Section 2.1).
This notations is concise and thus often convenient.

\subsection{Notions, notations, and assumptions on Boolean formulae}
\label{subsec:2_2}

In this subsection,
we describe notations and assumptions 
on Boolean formulae and satisfiability.
We assume the reader to be  familiar with them. 
The reader is referred to Chapter 4 in \cite{papadimitriou1995computational}
if necessary.
Let $X$ be a set of Boolean variables.
Let $\psi$ be a Boolean formula over $X$.
We denote by $\VAR(\psi)$
the set of all variables occurring in $\psi$.
We denote the set $X \cap \VAR(\psi)$ by $X_\psi$.
Let $\varphi$ be a CNF formula over $X$.
We denote by $\mathcal{C}_\varphi$ the set of all clauses in $\varphi$.
For any $x \in X$,
we denote its literal by $\tilde{x}$.
For any variable set $Z \subseteq X$,
we denote the set $\{\tilde{z} \colon z \in Z \}$ by $\tilde{Z}$.
For any literal $\tilde{x}$,
we denote its complement by $\tilde{x}^\mathrm{c}$.

Given a Boolean formula $\psi$,
we assume that there is a bijection from $\VAR(\psi)$
to $[1,|\VAR(\psi)|]$.
That is, all variables occurring in $\psi$
have successive indices from $1$.
If $\VAR(\varphi) = \{x_1,\cdots,x_n\}$
for a given Boolean formula $\psi$,
then we denote a truth assignment $\sigma$
by a binary representation $b_1 \cdots b_n$
such that $b_i = \sigma(x_i)$ for every $1 \leq i \leq n$.
For any assignment $\sigma$,
we denote by $T(\varphi, \sigma)$
the set of all literals $l$ in $\varphi$ such that $\sigma(l) = 1$.
For any assignment $\sigma$,
we say that a literal $l$ is {\itshape true} if $\sigma(l) = 1$;
and {\itshape false} otherwise.
We define the size of $\varphi$
as the number of occurrences of literals in $\varphi$.

Let $\psi$ be a Boolean formula over the union of
pairwise disjoint sets $X_1,\cdots,X_n$, where $n \geq 1$.
Following the above assumption,
we assume every variable in $\VAR(\varphi) \cap X_l$ 
to have an index in a suitable method
such that all variables in $\VAR(\varphi)$ have successive indices from $1$.
For example, for every $1 \leq l \leq n$,
we may consider every variable in $\VAR(\varphi) \cap X_l$ 
to have a positive integer in 
$[\sum_{i=1}^{l-1}|\VAR(\varphi) \cap X_i|+1, \sum_{i=1}^l|\VAR(\varphi) \cap X_i|]$
as its index.
Thus, we can denote an assignment for all variables in $\psi$
by $\sigma_1 \cdots \sigma_n$,
where $\sigma_i$ is a partial assignment for $\VAR(\psi) \cap X_i$
for every $1 \leq i \leq n$.

Let $\varphi$ be a CNF formula.
Throughout the paper,
we assume $|X_\varphi| \geq 2$ and $|\mathcal{C}_\varphi| \geq 2$.
Furthermore, for any variable $z$, 
every clause $C \in \mathcal{C}_\varphi$ includes at most one literal of $z$.
By these assumptions,
we do not lose the generality.

\subsection{Complexity classes and hierarchies}
\label{subsec:2_complexity_class}

In this paper, we assume that the reader is familiar with
basic notions and results
on the computational complexity theory.
The reader is referred to 
\cite{arora2009computational,du2011theory,papadimitriou1995computational}
if necessary.
We assume the alphabet of a given problem or relation to be $\{0,1\}$
unless stated otherwise.
Let $\mathcal{C}$ be a class of decision problems.
Let $L$ be a decision problem in $\mathcal{C}$.
We denote by $R_L$ a binary relation such that 
a pair $(u,v)$ is in $R_L$ if and only if
$v$ is a witness of $u$ in $L$.
We often call $R_L$ the {\itshape underlying binary relation} of $L$.
We assume $R_L$ to be polynomially balanced; i.e.,
for every pair $(u,v) \in R_L$, $|v| \leq p(|u|)$,
where $p$ is a polynomial.
For every instance $u$ of $L$,
we denote the set of all witnesses of $u$ by $W_L(u)$.
Let $f$ be a function from $\{0,1\}^\ast$ to $\mathbb{N}$
such that $f(u) = |W_L(u)|$ for a given $u \in \{0,1\}^\ast$.
We call $f$ the counting problem for $L$.
We often denote $f$ by $\NUM L$.
For any binary relation $R$,
we assume that every pair of $R$ is encoded over $\{0,1\}^\ast$.
We denote by $\NUM\DOT\mathcal{C}$
the class of all counting problems $\NUM S$ such that $R_S \in \mathcal{C}$.

We denote by $M^A$ an oracle Turing machine with an oracle $A$.
We denote the problem that an oracle machine $M^A$ accepts by $L(M^A)$.
We consider $M^\emptyset$ to be $M$.
Let $L^A$ be a problem $L$
such that $L(M^A) = L$ for some oracle machine $M^A$.
If $\mathcal{B}$ is a class of decision problems and 
$A$ is a decision problem, 
then we denote by $\mathcal{B}^A$ 
the class of all problems $B^A$ such that $B \in \mathcal{B}$.
If $\mathcal{A}$ is a class of decision problems,
then we denote by $\NUM \mathcal{A}$
the class of all counting problems $\NUM L$ such that
$L \in \NP^A$ for some $A \in \mathcal{A}$.
Note that for a given class $\mathcal{C}$ of decision problems,
if $co \mathcal{C}$ denotes the complement of $\mathcal{C}$,
then $\NUM co \mathcal{C} = \NUM\mathcal{C}$
by definition.
If $\mathcal{A}$ and $\mathcal{B}$ are classes of decision problems,
then we denote the class $\{\mathcal{B}^A \colon A \in \mathcal{A}\}$ 
by $\mathcal{B}^\mathcal{A}$.

We define the classes $\SIGMAP{k}$ and $\PIP{k}$,
where $k \geq 0$,
introduced by \cite{MeyerStockmeyer1972} inductively as follows.
$\SIGMAP{0}$ is the class $\CLASSP$.
For any $k \geq 1$, $\SIGMAP{k}$ is the class $\NP^\SIGMAP{k-1}$.
For every $k \geq 0$, $\PIP{k}$ is the class of the complements 
of all problems in $\SIGMAP{k}$. 
We call the sequence $\SIGMAP{0},\cdots,\SIGMAP{k},\cdots$ 
of the classes the polynomial hierarchy.
We denote the polynomial hierarchy by $\PH$.
By definition,
the classes $\SIGMAP{1}$ and $\PIP{1}$ coincide with
$\NP$ and $\CONP$, respectively.
The polynomial hierarchy has other characterizations.
We can characterize $\PH$
by polynomially balanced binary relations~\cite{Wrathall197623}.
Let $k \geq 1$.
A decision problem $L$ is in $\SIGMAP{k}$
if and only if its underlying binary relation $R_L$ is in $\PIP{k-1}$. 
The other characterization of $\PH$
is by $k$-alternating Turing machines~\cite{Chandra:1981:ALT:322234.322243}.
In this paper, we use those three characterizations.
Let $k \geq 0$.
We adopt a notation for decision problems in $\SIGMAP{k}$ and $\PIP{k}$.
Let $L$ be a decision problem.
We define decision problems 
$\SIGMA{k} L$ and $\PI{k} L$ inductively as follow.
$\SIGMA{1} L$ is $L$. 
For every $k \geq 2$, $\SIGMA{k} L$ is a decision problem $S$ 
such that its underlying binary relation $R_S$ is in $\PI{k-1} L$.
For every $k \geq 1$, $\PI{k} L$ is the complement of $\SIGMA{k} L$.

Given a decision problem in $\SIGMAP{k}$ or $\PIP{k}$, where $k \geq 0$,
another type of counting problem were studied
by Bauland et al.~\cite{BaulandBohlerCreignouReithSchnoorVollmer2010TOCS,Durand2005496}. 
Let $L$ be a decision problem.
We denote by $\NUMD{\SIGMA{k} L}$ and $\NUMD{\PI{k} L}$ 
the following counting problems $f_1$ and $f_2$, respectively. 
Given $w \in \{0,1\}^\ast$,
$f_1(w) = |W_K(w)|$ and $f_2(w) = |W_J(w)|$, 
where $K$ and $J$ are decision problems
such that $R_K \in \SIGMA{k} L$ and $R_J \in \PI{k} L$, respectively.
Note that the decision counterpart of $\NUMD{\PI{k-1} L}$
is $\SIGMA{k} L$
while the decision counterpart of $\NUM\SIGMA{k} L$
is $\SIGMA{k} L$.
This asymmetry is inherent in counting problems
and is from the task that counts the existing solutions for a given input.

By the definitions of the polynomial hierarchy 
and $\NUM\mathcal{C}$ and $\NUMD{\mathcal{C}}$ 
for some decision problem class $\mathcal{C}$,
we immediately obtain the following hierarchies of counting complexity classes.
We denote by $\NUM\PH$ and $\NUMD{\PH}$
the sequences of 
$\NUM\SIGMAP{0},\cdots,\NUM\SIGMAP{k},\cdots$
and $\NUMD{\SIGMAP{0}},\cdots,\NUMD{\SIGMAP{k}},\cdots$
of the counting complexity classes, respectively.

\subsection{Relaxed subtractive reductions and its special cases}
\label{subsec:2_reducibilities}

Durand et al.~\cite{Durand2005496} introduced subtractive reductions and 
found some complete problems on counting under this reductions.
We introduce a new type of reduction by generalizing subtractive reductions.

\begin{definition}
\label{def:4}
Let $A$ and $B$ be decision problems, respectively.
We define a {\itshape strong relaxed subtractive reduction} 
from $\NUM A$ to $\NUM B$ as a pair $(t_0,t_1)$,
where $t_0$ and $t_1$ are polynomial-time computable functions
from $\{0,1\}^\ast$ to $\{0,1\}^\ast$,
which satisfy the following.
There is a polynomially balanced binary relation $R_F \subset \{0,1\}^\ast \times \{0,1\}^\ast$ 
such that $\NUM F \in \FP$ and for any $w \in \{0,1\}^\ast$,

\begin{enumerate}
 \item $W_F(w) \subseteq W_B(t_0(w))$,
 \item $W_B(t_1(w)) \subseteq W_B(t_0(w))$,
 \item $W_B(t_1(w)) \cap W_F(w) = \emptyset$,
 \item $|W_A(w)| = |W_B(t_0(w))| - |W_B(t_1(w))| - |W_F(w)|$.
\end{enumerate}

\end{definition}

\begin{figure}[htbp]
\begin{center}
\includegraphics{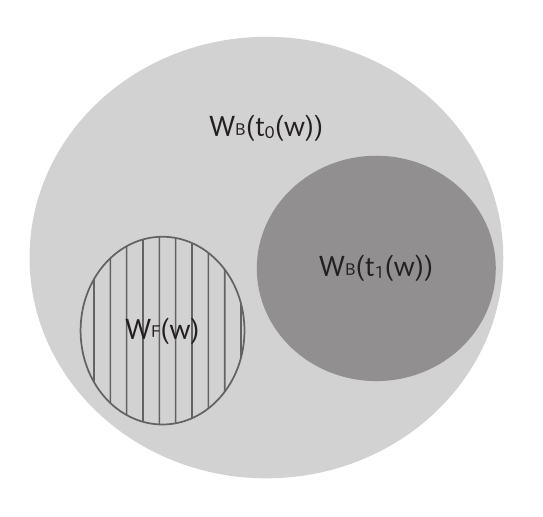}
\end{center}
\caption{Euler diagram for a strong relaxed subtractive reduction}
\label{fig:2}
\end{figure}

Figure~\ref{fig:2} illustrates an Euler diagram for three related sets in a strong relaxed subtractive reduction.
Let $K$ and $L$ be problems.
Let $\xi$ be a type of polynomial-time reduction.
Then, we say that $K$ is {\itshape $\xi$ reducible} to $L$
if there is a $\xi$ reduction from $K$ to $L$.
We say that a relation $R$ is the {\itshape $\xi$ reducibility}
if $R$ consists of all pairs $(K,L)$ such that
$K$ is $\xi$ reducible to $L$.

\begin{definition}
\label{def:2}
We call the reflexive and transitive relation of strong relaxed subtractive reducibility
the {\itshape relaxed subtractive reducibility};
i.e., we call a sequence $(t_{01},t_{11})$, $\cdots$, $(t_{0n},t_{1n})$
a {\itshape relaxed subtractive reduction}
if there is a sequence $A_0,\cdots,A_n$ of problems such that
$(t_{0i},t_{1i})$ is a strong relaxed subtractive reduction
from $A_{i-1}$ to $A_i$ for every $1 \leq i \leq n$.
\end{definition}

We define some special cases of (strong) relaxed subtractive reductions
as follows.
Let $A$, $B$, $(t_0,t_1)$, and $F$ be defined as in \ref{def:4}. 
We call $(t_0,t_1)$ a {\itshape strong subtractive reduction} 
if $W_F(w) = \emptyset$ for every $w \in \{0,1\}^\ast$.
Similarly, we define {\itshape subtractive reductions}.
We call $(t_0,t_1)$ a {\itshape parsimonious reduction} 
if $t_1(w) = W_F(w) = \emptyset$ for every $w \in \{0,1\}^\ast$.
We usually denote a parsimonious reduction $(t_0,t_1)$ by simply $t_0$.
Parsimonious reductions have the following significant property 
for discussing the completenesses for classes $\NUMD{\SIGMAP{k}}$ and $\NUMD{\PIP{k}}$.

\begin{theorem}[\cite{BaulandBohlerCreignouReithSchnoorVollmer2010TOCS,Durand2005496}]
The classes $\NUMD{\SIGMAP{k}}$ and $\NUMD{\PIP{k}}$ 
are closed under parsimonious reductions.
\end{theorem}

\subsection{Related computational problems}
\label{subsec:computational_problems}

For convenience,
we summarize the computational problems
that we describe in this paper.

\subsubsection{Decision problems}
\label{subsubsec:decision_problems}

\begin{problem}[$\GAP$] \ 

\noindent{\bfseries Input.} 
A set $A$ of coprime positive integers $a_1,\cdots,a_n$
such that $2 \leq a_1 < \cdots < a_n$ and $n \geq 2$.

\noindent{\bfseries Question.} 
Is there a gap of $\mathcal{S}(A)$?

\noindent{\bfseries Comment.}
This problem is trivial since the answer is always ``Yes''.

\end{problem}

\begin{problem}[$\TGAP$] \ 

\noindent{\bfseries Input.} 
A pair $(A, \kappa)$, 
where $A$ is a set of coprime positive integers $a_1,\cdots,a_n$
such that $2 \leq a_1 < \cdots < a_n$ and
$n \geq 2$, and $\kappa \in \mathbb{N}$.

\noindent{\bfseries Question.} 
Is there an integer in $N(A) \cap [\kappa, \infty]$? 

\end{problem}

\begin{problem}[$\IGAP$] \

\noindent{\bfseries Input.} 
A pair $(A, [\kappa_0,\kappa_1])$, 
where $A$ is a set of positive integers $a_1,\cdots,a_n$
such that $2 \leq a_1 < \cdots < a_n$ and
$n \geq 2$, and $\kappa_0, \kappa_1 \in \mathbb{N}$.

\noindent{\bfseries Question.} 
Is there an integer in $N(A) \cap [\kappa_0, \kappa_1]$? 

\noindent{\bfseries Comment.}
`` $\IGAP$'' is an abbreviation for ``nonrepresentable''.

\noindent{\bfseries Note.}
$a_1,\cdots,a_n$ may not necessarily be coprime.

\end{problem}

As a comprehensive survey of the decision problems in the $2$nd and $3$rd levels of the polynomial hierarchy
and the results on them,
we can refer articles due to Marcus Sch\"afer and Umans~\cite{schaefer2002completeness,schaefer2002completenessII}.

\subsubsection{Counting problems}
\label{subsubsec:counting_problems}

\begin{problem}[$\NUM\GAP$] \ 

\noindent{\bfseries Input.} 
A set $A$ of coprime positive integers $a_1,\cdots,a_n$
such that $2 \leq a_1 < \cdots < a_n$ and $n \geq 2$.

\noindent{\bfseries Output.} 
$|N(A)|$;
i.e., the number of gaps of $\mathcal{S}(A)$.

\end{problem}

\begin{problem}[$\NUM\TGAP$] \ 

\noindent{\bfseries Input.} 
A pair $(A, \kappa)$, 
where $A$ is a set of coprime positive integers $a_1,\cdots,a_n$
such that $2 \leq a_1 < \cdots < a_n$ and
$n \geq 2$, and $\kappa \in \mathbb{N}$.

\noindent{\bfseries Output.} 
$|N(A) \cap [\kappa, \infty]|$; i.e.,
the number of gaps of $\mathcal{S}(A)$,
which is greater than or equal to $\kappa$.

\end{problem}

\begin{problem}[$\NUM\IGAP$] \

\noindent{\bfseries Input.} 
A pair $(A, [\kappa_0,\kappa_1])$, 
where $A$ is a set of positive integers $a_1,\cdots,a_n$
such that $2 \leq a_1 < \cdots < a_n$ and
$n \geq 2$, and $\kappa_0, \kappa_1 \in \mathbb{N}$.

\noindent{\bfseries Output.} 
$|N(A) \cap [\kappa_0,\kappa_1]|$; i.e.,
the number of gaps of $\mathcal{S}(A)$ in $[\kappa_0,\kappa_1]$.

\noindent{\bfseries Comment.}
`` $\IGAP$'' is an abbreviation for ``nonrepresentable''.

\noindent{\bfseries Note.}
$a_1,\cdots,a_n$ may not necessarily be coprime.

\end{problem}

To formulate related counting problems on Boolean satisfiability,
we define two sets $\Phi$ and $\Phi_{1/3}$ as follows. 

\begin{definition}
\label{def:1}
$\Phi$ is the set of all triples $(k,\varphi, \sigma)$,
where $k$ is a positive integer, 
$\varphi$ is a CNF-formula $C_1 \land \cdots \land C_m$,
and $\sigma$ is a truth assignment,
such that $(k,\varphi, \sigma) \in \Phi$
if and only if $k$ is even and $|C_\alpha \cap T(\varphi, \sigma)| \neq 0$ 
for every $1 \leq \alpha \leq m$ 
or $k$ is odd and $|C_\alpha \cap T(\varphi, \sigma)| = 0$ 
for some $1 \leq \alpha \leq m$.
\end{definition}

\begin{definition}
\label{def:3}
$\Phi_{1/3}$ is the set of all triples $(k,\varphi, \sigma)$,
where $k$ is a positive integer, 
$\varphi$ is a CNF-formula $C_1 \land \cdots \land C_m$,
and $\sigma$ is a truth assignment, 
such that $(k,\varphi, \sigma) \in \Phi_{1/3}$ 
if and only if 
$k$ is even and $|C_\alpha \cap T(\varphi, \sigma)| = 1$ 
for every $1 \leq \alpha \leq m$ 
or $k$ is odd and $|C_\alpha \cap T(\varphi, \sigma)| \neq 1$ 
for some $1 \leq \alpha \leq m$.
\end{definition}

\begin{problem}[$\NUMD{\SIGMA{k}\SAT}$, $k \geq 0$] \

\noindent{\bfseries Input.} 
A CNF formula $\varphi$ over $X \cup \bigcup_{1 \leq i \leq k} X_i$,
where $X, X_1,\cdots,X_k$ are pairwise disjoint.

\noindent{\bfseries Output.} 
$|\{\sigma \in \{0,1\}^{|X_\varphi|} \colon
\exists \sigma_1 \forall \sigma_2 \cdots Q \sigma_k 
[(k,\varphi, \sigma \sigma_1 \cdots \sigma_k) \in \Phi] \}|$,
where 
$\sigma_i$ is a partial assignment for $X_{i,\varphi}$ for every $i$ with $1 \leq i \leq k$.

\noindent{\bfseries Comment.}
It is known to be $\NUMD{\SIGMAP{k}}$-complete under parsimonious reductions~\cite{Durand2005496}.

\noindent{\bfseries Note.}
We consider the problem $\NUMD{\SIGMA{0}\SAT}$ 
to be $\NUM\SAT$.

\end{problem}

\begin{problem}[$\NUMD{\PI{k}\SAT}$, $k \geq 1$] \

\noindent{\bfseries Input.} 
A CNF formula $\varphi$ over $X \cup \bigcup_{1 \leq i \leq k} X_i$,
where $X, X_1,\cdots,X_k$ are pairwise disjoint.

\noindent{\bfseries Output.} 
$|\{\sigma \in \{0,1\}^{|X_\varphi|} \colon
\forall \sigma_1 \exists \sigma_2 \cdots Q \sigma_k
[(k,\varphi, \sigma \sigma_1 \cdots \sigma_k) \not\in \Phi]
\}|$,
where 
$\sigma_i$ is a partial assignment for $X_{i,\varphi}$ for every $i$ with $1 \leq i \leq k$.

\noindent{\bfseries Comment.}
It is known to be $\NUMD{\PIP{k}}$-complete under parsimonious reductions~\cite{Durand2005496}.

\end{problem}

\begin{problem}[$\NUMD{\SIGMA{k}\OT\SAT}$, $k \geq 0$] \

\noindent{\bfseries Input.} 
A $3$-CNF formula $\varphi$ over $X \cup \bigcup_{1 \leq i \leq k} X_i$,
where $X, X_1,\cdots,X_k$ are pairwise disjoint.

\noindent{\bfseries Output.} 
$|\{\sigma \in \{0,1\}^{|X_\varphi|} \colon
\exists \sigma_1 \forall \sigma_2 \cdots Q \sigma_k
[(k,\varphi, \sigma \sigma_1 \cdots \sigma_k) \in \Phi_{1/3}] \}|$,
where 
$\sigma_i$ is a partial assignment for $X_{i,\varphi}$ for every $i$ with $1 \leq i \leq k$.

\noindent{\bfseries Note.}
We consider the problem $\NUMD{\SIGMA{0}\OT\SAT}$ to be $\NUM\OT\SAT$.

\end{problem}

\begin{problem}[$\NUMD{\PI{k}\OT\SAT}$, $k \geq 1$] \

\noindent{\bfseries Input.} 
A $3$-CNF formula $\varphi$ over $X \cup \bigcup_{1 \leq i \leq k} X_i$,
where $X, X_1,\cdots,X_k$ are pairwise disjoint.

\noindent{\bfseries Output.} 
$|\{\sigma \in \{0,1\}^{|X_\varphi|} \colon
\forall \sigma_1 \exists \sigma_2 \cdots Q \sigma_k
[(k,\varphi, \sigma \sigma_1 \cdots \sigma_k) \not\in \Phi_{1/3}] \}|$,
where 
$\sigma_i$ is a partial assignment for $X_{i,\varphi}$ for every $i$ with $1 \leq i \leq k$.

\end{problem}

\begin{example}
\label{ex:2}

Let $\varphi_1$ be a $3$-CNF formula $C_1 \land C_2 \land C_3 \land C_4$
over $X \cup Y$,
where $X$ and $Y$ are pairwise disjoint,
$C_1 = \lnot \lor x_1 \lor x_2 \lor  y_1$,
$C_2 = x_1 \lor x_3 \lor \lnot y_2$,
$C_3 = \lnot x_2 \lor x_4 \lor \lnot y_3$,
$C_4 = x_4 \lor y_2 \lor y_3$,
$X_{\varphi_1} = \{x_1,x_2,x_3,x_4\}$
and $Y_{\varphi_1} = \{y_1,y_2,y_3\}$.
We can check the values of every assignment in Table~\ref{tab:2}.
In this table,
every cell corresponds to an assignment $\sigma$ for $\varphi_1$
and contains the indices of all clauses $C$ such that
$|T(C,\sigma)|=1$.
In the case where $\sigma_x = 0000$, 
if we let $\sigma_y$ be $001$, then
exactly one literal is $1$ in every clause.
Similarly, in the case where $\sigma_x = 1100$, 
if we let $\sigma_y$ be $010$, 
then exactly one literal is $1$ in every clause.
However, if $\sigma_x$ is one of the other $X$-assignments,
no $Y$-assignment exists such that 
exactly one literal is $1$ in every clause.
Thus, by Table~\ref{tab:2}, 
we can observe that 
$\NUMD{\PI{1}\OT\SAT(\varphi_1)}$ is $14$;
and moreover, $\varphi_1$ is an yes-instance of $\SIGMA{2}\OT\SAT$.

\begin{table}[htbp]
\caption{Clauses in $\varphi_1$, each of which contains exactly one true literal,
where each cell corresponds to an assignment and contains indices of clauses.}
\label{tab:2}
\begin{tabular}{ccccccccc}
\hline
\multirow{2}{*}{$\sigma_x$} & \multicolumn{8}{c}{$\sigma_y$} \\
\cline{2-9} 
 & $000$ & $100$ & $010$ & $001$ & $110$ & $101$ & $011$ & $111$ \\
\hline
\rowcolor{MyGray} $0000$ & {$1,2$} &  {$2$} &  {$1,4$} &  {$1,2,3,4$} &  {$4$} &  {$2,3,4$} &  {$1,3$} &  {$3$} \\
 $1000$ & --- & {$1$} &  {$2,4$} &  {$3,4$} &  {$1,2,4$} &  {$1,3,4$} &  {$2,3$} &  {$1,2,3$} \\
 $0100$ & {$2,3$} &  {$2,3$} &  {$3,4$} &  {$2,4$} &  {$3,4$} &  {$2,4$} &  --- & --- \\
 $0010$ & {$1$} &  --- & {$1,2,4$} &  {$1,3,4$} &  {$2,4$} &  {$3,4$} &  {$1,2,3$} &  {$2,3$} \\
 $0001$ & {$1,2,4$} &  {$2,4$} &  {$1$} &  {$1,2$} &  --- & {$2$} &  {$1$} &  --- \\
\rowcolor{MyGray} $1100$ & {$1,3$} &  {$3$} &  {$1,2,3,4$} &  {$1,4$} &  {$2,3,4$} &  {$4$} &  {$1,2$} &  {$2$} \\
 $1010$ & --- & {$1$} &  {$4$} &  {$3,4$} &  {$1,4$} &  {$1,3,4$} &  {$3$} &  {$1,3$} \\
 $1001$ & {$4$} &  {$1,4$} &  {$2$} &  --- & {$1,2$} &  {$1$} &  {$2$} &  {$1,2$} \\
 $0110$ & {$3$} &  {$3$} &  {$2,3,4$} &  {$4$} &  {$2,3,4$} &  {$4$} &  {$2$} &  {$2$} \\
 $0101$ & {$2,4$} &  {$2,4$} &  --- & {$2,3$} &  --- & {$2,3$} &  {$3$} &  {$3$} \\
 $0011$ & {$1,4$} &  {$4$} &  {$1,2$} &  {$1$} &  {$2$} &  --- & {$1,2$} &  {$2$} \\
 $1110$ & {$1,3$} &  {$3$} &  {$1,3,4$} &  {$1,4$} &  {$3,4$} &  {$4$} &  {$1$} &  --- \\
 $1101$ & {$1,4$} &  {$4$} &  {$1,2$} &  {$1,3$} &  {$2$} &  {$3$} &  {$1,2,3$} &  {$2,3$} \\
 $1011$ & {$4$} &  {$1,4$} &  --- & --- & {$1$} &  {$1$} &  --- & {$1$} \\
 $0111$ & {$4$} &  {$4$} &  {$2$} &  {$3$} &  {$2$} &  {$3$} &  {$2,3$} &  {$2,3$} \\
 $1111$ & {$1,4$} &  {$4$} &  {$1$} &  {$1,3$} &  --- & {$3$} &  {$1,3$} &  {$3$} \\
\hline
\end{tabular}
\end{table}

\end{example}

We summarize the relations between decision problems and their counting problems
and their complexity classes in Table~\ref{tab:related_problems_complexities}.

\begin{table}[htbp]
\begin{center}
\caption{Complexities of related computational problems.}
\label{tab:related_problems_complexities}
\begin{tabular}{p{10em}p{9em}p{9em}}
\hline
Counting & Upper  & Decision
\\
problems & bounds & problems
\\
& [reduction type] & 
\\
\cline{1-3}
\multirow{2}{*}{$\NUM\IGAP$} & \cellcolor{MyGrayLight}{$\NUM\NP$-complete} & \\
& \cellcolor{MyGrayLight}{[parsimonious] \ \ \ \ \ \ \ \ (Section~\ref{sec:proof_parsimonious_reducibility})} & $\IGAP$\\
\cline{1-3}
\multirow{2}{*}{$\NUM\TGAP$} & \cellcolor{MyGrayLight}{$\NUM\NP$-complete} & \\
& \cellcolor{MyGrayLight}{[parsimonious] \ \ \ \ \ \ \ \ (Section~\ref{sec:proof_bounded_gaps_parsimonious_reducibility})} & $\TGAP$ \\
\cline{1-3}
\multirow{2}{*}{$\NUM\GAP$} & \cellcolor{MyGrayLight}{$\NUM\NP$-complete} & \\
& \cellcolor{MyGrayLight}{[relaxed subtractive] \ \ \ \ \ \ \ \ (Section~\ref{sec:proof_relax_subtractive_reducibility})} & $\GAP$ \\
\hline
\multirow{3}{*}{$\NUM \SAT$} & $\NUM \CLASSP$-complete 
& \multirow{6}{*}{$\SAT$} \\
& [parsimonious] \cite{VALIANT1979TCS} & \\
\cline{1-2}
\multirow{3}{*}{$\NUMD{\SIGMA{1}\SAT}$} & $\NUM\NP$-complete & \\
& [parsimonious] & \\
& \cite{Durand2005496} &
\\
\cline{1-3}
\multirow{3}{*}{$\NUMD{\PI{k-1}\SAT}$} & $\NUMD{\PIP{k-1}}$-complete 
& \multirow{6}{*}{$\SIGMA{k}\SAT$} \\
& [parsimonious] & \\
& \cite{Durand2005496}  &
\\
\cline{1-2}
\multirow{3}{*}{$\NUMD{\SIGMA{k}\SAT}$} & $\NUMD{\SIGMAP{k}}$-complete & \\
& [parsimonious] & \\
& \cite{Durand2005496} & 
\\
\cline{1-3}
\multirow{2}{*}{$\NUMD{\OT\SAT}$} & $\NUM\CLASSP$-complete
& \multirow{4}{*}{$\OT\SAT$} \\
& [parsimonious] &
\\
\cline{1-2}
\multirow{2}{*}{$\NUMD{\SIGMA{1}\OT\SAT}$} & $\NUM\NP$-complete & \\
& [parsimonious] &
\\
\cline{1-3}
\multirow{2}{*}{$\NUMD{\PI{k-1}\OT\SAT}$} & $\NUMD{\PIP{k-1}}$-complete
& \multirow{4}{*}{$\SIGMA{k}\OT\SAT$} \\
& [parsimonious] & \\
\cline{1-2}
\multirow{2}{*}{$\NUMD{\SIGMA{k}\OT\SAT}$} & $\NUMD{\SIGMAP{k}}$-complete & \\
& [parsimonious] &
\\
\hline
\end{tabular}
\end{center}
\end{table}

\subsection{The complexities of $\NUM\SIGMA{k}\OT\SAT$ and $\NUM\PI{k}\OT\SAT$}
\label{subsec:2_sat_completeness}

In this section, we describe 
the $\NUM\DOT\PIP{k}$-completeness of $\NUM\PI{k}\OT\SAT$.
This statement is equivalent to
Lemma 5.5 in \cite{BaulandBohlerCreignouReithSchnoorVollmer2010TOCS}
and they proved it by using a theory 
for constraints satisfaction problems, 
which is a more general tool for investigating
satisfiability problems than 
the one in this paper.
For readability,
we give a slightly weaker but more intuitive and elementary proof.
The reduction in the following proof is a folklore,
although any explicit literature has not been published,
to the best of the author's knowledge.

\begin{theorem}
\label{thm:6}
$\NUMD{\SIGMA{k}\OT\SAT}$ is $\NUM\DOT\SIGMAP{k}$-complete under parsimonious reductions
for every $k \geq 0$.
\end{theorem}

\noindent {\itshape Proof.}
By Theorem 2.1~in~\cite{Durand2005496},
$\NUMD{\SIGMA{k}\SAT}$ is $\NUMD{\SIGMAP{k}}$-complete under parsimonious reductions.
Since $\NUMD{\SIGMA{k}\OT\SAT}$ is a special case of $\NUMD{\SIGMA{k}\SAT}$,
the problem $\NUMD{\SIGMA{k}\OT\SAT}$ is in $\NUMD{\SIGMAP{k}}$.
Thus, it suffices to show that $\NUMD{\SIGMA{k}\SAT}$ is parsimonious reducible 
to $\NUMD{\SIGMA{k}\OT\SAT}$.
Let $X, Y_1,\cdots,Y_{k-1}$ be pairwise disjoint sets of variables.
Let $Z$ be the set $X \cup \bigcup_{i=1}^{k-1} Y_i$.
Let $\varphi$ be a $3$-CNF formula over $Z$.
We will define a set $V$ of new variables 
and then construct a $3$-CNF formula $\varphi^\prime$
such that
\begin{eqnarray*}
&& |\{\sigma^\prime \in \{0,1\}^{|X_\varphi \cup V|} \colon
\exists \sigma_1^\prime \forall \sigma_2^\prime \cdots Q \sigma_{k-1}^\prime
[(k,\varphi^\prime, \sigma^\prime \sigma_1^\prime \cdots \sigma_{k-1}^\prime) \in \Phi_{1/3}] \}|
\\
&& =|\{\sigma \in \{0,1\}^{|X_\varphi|} \colon
\exists \sigma_1 \forall \sigma_2 \cdots Q \sigma_{k-1}
[(k,\varphi, \sigma \sigma_1 \cdots \sigma_{k-1}) \in \Phi] \}|,
\end{eqnarray*}
where $\sigma_i$ and $\sigma_i^\prime$ denote partial assignments for $X_i$
for every $i$ with $1 \leq i \leq k-1$.

Let $C$ be a clause $\tilde{z}_1 \lor \tilde{z}_2 \lor \tilde{z}_3$ in $\varphi$,
where $\tilde{z}_1, \tilde{z}_2$, and $\tilde{z}_3$ are literals over $V \cup Z$.
Let $\langle C \rangle_1, \cdots, \langle C \rangle_9$ be new variables.
Then we define a $3$-CNF formula $\varphi_C^\prime$ as the conjunction of clauses
\begin{eqnarray*}
&& \tilde{z}_1 \lor \langle C\rangle_1 \lor \langle C \rangle_2, \ \
\tilde{z}_2 \lor \langle C \rangle_2 \lor \langle C \rangle_3, \ \ 
\tilde{z}_3 \lor \langle C\rangle_5 \lor \langle C\rangle_6,
\\
&& \tilde{z}_3 \lor \langle C \rangle_6 \lor \langle C \rangle_7, \ \ 
\langle C \rangle_1 \lor \langle C \rangle_3 \lor \langle C \rangle_4,  \ \
\langle C \rangle_5 \lor \langle C \rangle_7 \lor \langle C \rangle_8,
\\
&& 
\langle C \rangle_2 \lor \langle C \rangle_6 \lor \langle C \rangle_9. \ \ 
\end{eqnarray*}
Table~\ref{tab:in_proof_1} illustrates
all satisfying assignments for $\varphi^\prime_C$. 
We define $V$ as 
the set 
\begin{eqnarray*}
\{\langle C \rangle_i \colon C \text{\ is a clause in } \varphi, 1 \leq i \leq 9 \}.
\end{eqnarray*}
Finally, we define $\varphi^\prime$
as the conjunction of $\varphi_C^\prime$ for all clauses $C$ in $\varphi$.

We can check the following statement.
For every assignments $\sigma \in \{0,1\}^{|X_\varphi|}$
and $\delta \in \{0,1\}^{|Z|}$,
if $(k,\varphi, \sigma \delta) \in \Phi$ 
then
there is exactly one $\sigma^\prime \in \{0,1\}^{|X_\varphi \cup V|}$
such that
$(k,\varphi, \sigma^\prime \delta) \in \Phi_{1/3}$.
Moreover, we can check the statement of the converse direction.
We can construct $\varphi^\prime$ in polynomial time
since we construct
$9$ variables and a formula of $7$ clauses 
for every clause in a given $\varphi$.
\hfill $\square$

\begin{table}
 \caption{All satisfying assignments for the constructed formula $\varphi^\prime_C$ from a clause
in the proof for Theorem~\ref{thm:6}.}
\label{tab:in_proof_1}
\begin{tabular}{cccccccccccccccccc}
\hline
$\tilde{z}_1$ & $\tilde{z}_2$ & $\tilde{z}_3$ &
$\langle C \rangle_1$ & $\langle C \rangle_2$ & 
$\langle C \rangle_3$ & $\langle C \rangle_4$ & 
$\langle C \rangle_5$ & $\langle C \rangle_6$ & 
$\langle C \rangle_7$ & $\langle C \rangle_8$ & 
$\langle C \rangle_9$ & \\
\hline
$0$ & $0$ & $1$ & $0$ & $1$ & $0$ & $1$ & $0$ & $0$ & $0$ & $1$ & $0$ 
\\
$0$ & $1$ & $0$ & $1$ & $0$ & $0$ &  $0$ & $0$ & $1$ & $0$ & $1$ & $0$ \\
$0$ & $1$ & $1$ & $1$ & $0$ & $0$ & $0$ & $0$ & $0$ & $0$ & $1$ & $1$ \\
$1$ & $0$ & $0$ & $0$ & $0$ & $1$ & $0$ & $0$ & $1$ & $0$ & $1$ & $0$ \\
$1$ & $0$ & $1$ & $0$ & $0$ & $1$ & $0$ & $0$ & $0$ & $0$ & $1$ & $1$ \\
$1$ & $1$ & $0$ & $0$ & $0$ & $0$ & $1$ & $0$ & $1$ & $0$ & $1$ & $0$ \\
$1$ & $1$ & $1$ & $0$ & $0$ & $0$ & $1$ & $0$ & $0$ & $0$ & $1$ & $1$ \\
\hline
\end{tabular}
\end{table}

Durand et al.\cite{Durand2005496} describes
the $\NUM\DOT\SIGMAP{k}$-completeness under parsimonious reductions
explicitly without proofs.
As they described in the paper~\cite{Durand2005496},
the proof for the statement is due to \cite{Wrathall197623}.
Although the proof in \cite{Wrathall197623}
is for the $\SIGMAP{k}$-completenesses of $\SIGMA{k}\SAT$,
the reduction is parsimonious.
Theorem 3.13 in \cite{du2011theory} also 
give a more concise proof for the same statement.

By the proof for Theorem~\ref{thm:6}, we immediately obtain the following corollary.
\begin{corollary}
\label{cor:1} 
$\NUMD{\PI{k}\OT\SAT}$ is $\NUM\DOT\PIP{k}$-complete under parsimonious reductions
for every $k \geq 1$.
\end{corollary}

\section{Overview of three versions of the numerical-semigroup-gap counting problem}
\label{sec:overview_parsimonious_reducibility}

By definition,
$\NUM\IGAP$ is a generalization of $\NUM\TGAP$ 
and $\NUM\TGAP$ is a generalization of $\NUM\GAP$. 
We can observe these relationships as follows.
Given a set $A$ of coprime positive integers $a_1,\cdots,a_n$ 
with $2 \leq a_1 < \cdots < a_n$ and $n \geq 2$,
the Frobenius number $g(A)$
is known to be smaller than $a_n^2$~\cite{Wilf1978}.
Since $g(A)$ is the largest element of $N(A)$
and $N(A)$ consists of only positive integers,
the cardinality of $N(A)$ is less than $a_n^2$.
For example, 
if $a_1=12, a_2 = 19, a_3 = 51, a_4 = 53$ and $A=\{a_1,a_2,a_3,a_4\}$, 
then $|N(A)| = 60$, $g(A) = 109$, and $a_4^2 = 2809$.
Thus, 
we consider an instance $(A,\lambda)$ of $\NUM\TGAP$
to be an instance $(A,[\lambda,a_n^2])$ of $\NUM\IGAP$;
and an instance $A$ of $\NUM\GAP$
to be an instance $(A,1)$ of $\NUM\TGAP$.
Thus, the hardnesses of three problems are 
\begin{center}
$\NUM\GAP$, \ \ \  $\NUM\TGAP$, \ \ \ $\NUM\IGAP$
\end{center}
in the order from the weakest one.

On the other hand, 
the difficulties of reductions from $\NUMD{\PI{1}\OT\SAT}$
appear to be in the reverse order for the ones of the problems in themselves.
$\NUM\IGAP$ has a structure that is the most similar 
to $\NUMD{\PI{1}\OT\SAT}$ among the three problems. 
$\NUM\GAP$ has a structure that is the farthest to $\NUMD{\PI{1}\OT\SAT}$.
We remark the following elementary fact.
We use \ref{rem:2}
in Sections \ref{sec:proof_bounded_gaps_parsimonious_reducibility} 
and \ref{sec:proof_relax_subtractive_reducibility},
implicitly.
\begin{remark}
\label{rem:2}
If $A_1$ and $A_2$ are sets of coprime positive integers such that $A_1 \subseteq A_2$, 
then $N(A_2) \subseteq N(A_1)$.
\end{remark}

To prove the $\NUM\NP$-completenesses of three problems $\NUM\IGAP$, $\NUM\TGAP$, and $\NUM\GAP$,
we use a standard approach for problems on numbers; e.g.,
the proofs for Theorem 3.5 in \cite{garey1979computers}
and Theorem 9.10 in \cite{papadimitriou1995computational}.
In Section~\ref{sec:proof_parsimonious_reducibility},
we will describe almost all parts of the reduction.
In Sections \ref{sec:proof_bounded_gaps_parsimonious_reducibility}%
~and~\ref{sec:proof_relax_subtractive_reducibility},
we will extend the reductions of the previous sections.

\section{Complexity of $\NUM \IGAP$}
\label{sec:proof_parsimonious_reducibility}

In this section, we investigate 
the complexity of $\NUM\IGAP$.
In particular, we prove the following theorem.
\begin{theorem}
\label{thm:2}
$\NUM\IGAP$ is $\NUM\NP$-complete under parsimonious reductions.
\end{theorem}

\subsection{Ideas of the proof}
\label{subsec:idea_reduction_to_gap_counting_problem_with_intervals}

In this subsection, 
we describe the idea of the proof for Theorem~\ref{thm:2}.
We can easily check the membership of $\NUM\IGAP$ to $\NUM\NP$.
Thus, the main part of the proof is for the hardness of $\NUM\IGAP$. 
The proof is by constructing a relaxed subtractive reduction
from $\NUMD{\PI{1}\OT\SAT}$ to $\NUM\IGAP$
in Section~\ref{subsec:proof_completeness_IGAP}.
We first describe the reduction intuitively and illustrate an example.

Let $\varphi$ be a $3$-CNF formula over $X \cup Y$,
where $X$ and $Y$ are pairwise disjoint.
To simulate the behavior of $\varphi$,
we define a positive integer for each pair of a truth assignment and a Boolean variable.
Let us describe the binary representations for integers that we construct.
Let $\beta$ be one of such binary representations.
We consider $\beta$ to be partitioned into $8$ {\itshape zones}.
Each of $8$ zones is classified into one of $5$ {\itshape types} with its role.
We explain each of the $5$ types in the order from the smaller bits.
The $1$st type is for simulating an assignment for a variable in $X$.
The $2$nd and $3$rd types are for simulating
which variable in $X$ and $Y$ we assign a Boolean value to, respectively.
The $4$th type is for simulating 
which clause we assign $1$ to a literal in.
The $5$th type is introduced for a purpose not directly for 
simulating anything in an assignment for $\varphi$
which we will describe below in more detail.

\begin{figure}[htbp]
\begin{center}
\includegraphics{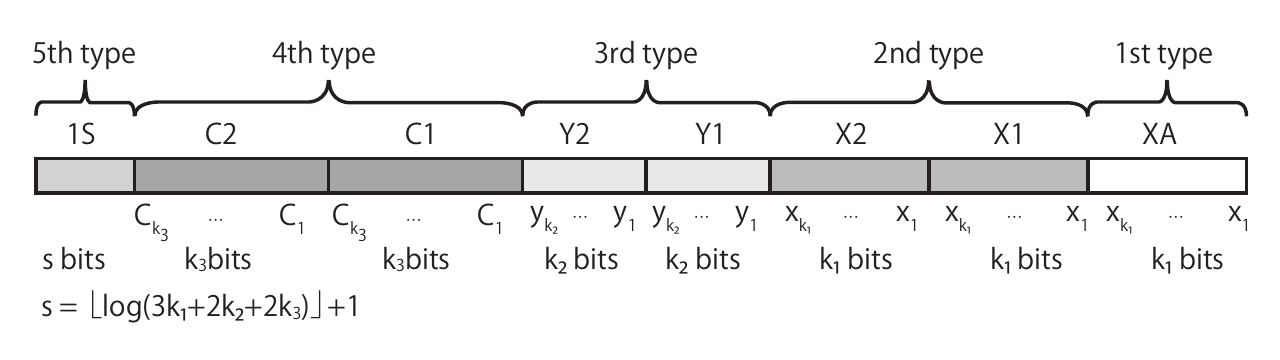}
\end{center}
\caption{Binary representation of a constructed integer from a $3$-CNF formula.}
\label{fig:structure_of_constructed_integer}
\end{figure}

Each type of subrepresentation is allocated 
to a fixed range of positions, which depends on $\varphi$.
Figure~\ref{fig:structure_of_constructed_integer} illustrates
the binary representation of a constructed integer.
In Figure~\ref{fig:structure_of_constructed_integer},
each of the $5$ types of subrepresentations is distinguished by a kind of shading of gray.
For a technical reason, the $2$nd to $4$th types of subrepresentations
have copies of them.
Such copies are 
not for the proof for Theorem~\ref{thm:2}
but for the ones for Theorems~\ref{thm:3}~and~\ref{thm:4}. 
We give each of $8$ zones a name as in Figure~\ref{fig:structure_of_constructed_integer}.
That is, the names of zones are
$\OSPT$, $\CTPT$, $\COPT$, $\YTPT$, $\YOPT$, $\XTPT$, $\XOPT$, and $\XAPT$
in the descending order from the highest bits.

We describe more details of integers that we construct.
Let $x_1,\cdots,x_{k_1}$ and $y_1,\cdots,y_{k_2}$
be all variables in $X$ and $Y$, occurring in $\varphi$, respectively.
Let $C_1,\cdots,C_{k_3}$ be all clauses in $\varphi$.
Every constructed integer is represented with 
$3k_1 + 2k_2 + 2k_3 + \lfloor \log(3k_1 + 2k_2 + 2k_3) \rfloor + 1$ bits,
where $k_1 = |X_\varphi|$, $k_2 = |Y_\varphi|$, and $k_3 = |\mathcal{C}_\varphi|$. 
Each of the $\XTPT$, $\XOPT$, and $\XAPT$ zones is of length $k_1$.
Each of the $\YTPT$ and $\YOPT$ zones is of length $k_2$.
Each of the $\CTPT$ and $\COPT$ zones is of length $k_3$.
For example, we construct integers for $\varphi_1$ in \ref{ex:2}
as illustrated in Table~\ref{tab:constructed_integers_from_varphi_one}.
In Table~\ref{tab:constructed_integers_from_varphi_one},
we denote the $\OSPT$ zone with its decimal representation
and the other zones with their binary representations
for reflecting the role of each zone.

\begin{table}[htbp]
\caption{Integers constructed from $\varphi_1 = C_1 \land C_2 \land C_3 \land C_4$, where 
$C_1 = \lnot x_1 \lor x_2 \lor y_1$, $C_2 = x_1 \lor x_3 \lor \lnot y_2$,
$C_3 = \lnot x_2 \lor x_4 \lor \lnot y_3$, $C_4 = x_4 \lor y_2 \lor y_3$.}
\label{tab:constructed_integers_from_varphi_one}
\begin{center}
\begin{tabular}{cc|cccccccc}
\hline
\multirow{2}{*}{Variables} & \multirow{2}{*}{Values}
& \multicolumn{7}{c}{Constructed integers} & \\
\cline{3-10}
& & $\OSPT$ & $\CTPT$ & $\COPT$ 
& $\YTPT$ & $\YOPT$ & $\XTPT$ & $\XOPT$ & $\XAPT$
\\
\hline
$x_1$ & $1$ & $5$ & $0010$ & $0010$ & $000$ & $000$ & $0001$ & $0001$ & $0001$ \\
$x_1$ & $0$ & $5$ & $0001$ & $0001$ & $000$ & $000$ & $0001$ & $0001$ & $0000$ \\
$x_2$ & $1$ & $5$ & $0001$ & $0001$ & $000$ & $000$ & $0010$ & $0010$ & $0010$ \\
$x_2$ & $0$ & $5$ & $0100$ & $0100$ & $000$ & $000$ & $0010$ & $0010$ & $0000$ \\
$x_3$ & $1$ & $5$ & $0010$ & $0010$ & $000$ & $000$ & $0100$ & $0100$ & $0100$ \\
$x_3$ & $0$ & $3$ & $0000$ & $0000$ & $000$ & $000$ & $0100$ & $0100$ & $0000$ \\
$x_4$ & $1$ & $7$ & $1100$ & $1100$ & $000$ & $000$ & $1000$ & $1000$ & $1000$ \\
$x_4$ & $0$ & $3$ & $1100$ & $1100$ & $000$ & $000$ & $1000$ & $1000$ & $0000$ \\
$y_1$ & $1$ & $4$ & $0001$ & $0001$ & $001$ & $001$ & $0000$ & $0000$ & $0000$ \\
$y_1$ & $0$ & $2$ & $0000$ & $0000$ & $001$ & $001$ & $0000$ & $0000$ & $0000$ \\
$y_2$ & $1$ & $4$ & $1000$ & $1000$ & $010$ & $010$ & $0000$ & $0000$ & $0000$ \\
$y_2$ & $0$ & $4$ & $0010$ & $0010$ & $010$ & $010$ & $0000$ & $0000$ & $0000$ \\
$y_3$ & $1$ & $4$ & $1000$ & $1000$ & $100$ & $100$ & $0000$ & $0000$ & $0000$ \\
$y_3$ & $0$ & $4$ & $0100$ & $0100$ & $100$ & $100$ & $0000$ & $0000$ & $0000$ \\
\hline
\end{tabular}
\end{center}
\end{table}

We describe the detail of the role of each zone of a binary representation that we construct.
A constructed integer corresponds to a pair of a Boolean variable and value.
Let $\beta$ be a binary representation that we construct.

Let us first consider $\beta$ to be one for simulating $x_i \in X$ assigned $1$, 
where $1 \leq i \leq k_1$.
In the $\XAPT$ zone, we set the $i$-th bit to $1$ and the other bits to $0$s.
This means that $x_i$ is assigned $1$ and the other variables are assigned $0$.
In each of the $\XOPT$ and $\XTPT$ zones,
we set the $i$-th bit to $1$ and the other bits to $0$s.
This means that $x_i$ is assigned a Boolean value
but the other variables are not assigned. 
In each of the $\YOPT$ and $\YTPT$ zones,
we set every bit to $0$.
This means that $b$ is not for simulating the behavior of a variable in $Y$.
Moreover, 
in each of the $\COPT$ and $\CTPT$ zones,
for every $1 \leq j \leq k_3$,
we set the $j$-th bit to $1$
if the clause $C_j$ includes the literal $x_i$;
and the other bits to $0$s otherwise.
Let $v$ be the binary representation of the integer $3 + 2l$
such that $l$ is the numbers of $1$s, occurring in the $\COPT$ zone.
We allocate $v$ to the $\OSPT$ zone of $\beta$.

Let us next consider $\beta$ to be one for simulating $x_i \in X$ assigned $0$.
We set every bit of the $\XAPT$ zone to $0$.
This means that at least $x_i$ is assigned $1$,
but does not mean that each of the remaining variables is assigned $0$.
The $\XOPT$, $\XTPT$, $\YOPT$, and $\YTPT$ zones 
are constructed in the same way as for the case where $x_i$ is assigned $1$.
In each of the $\COPT$ and $\CTPT$ zones,
for every $1 \leq j \leq k_3$,
we set the $j$-th bit to $1$
if the clause $C_j$ includes the literal $\lnot x_i$;
and the other bits to $0$s otherwise.
Let $v$ be the binary representation of the integer $2 + 2l$
such that $l$ is the numbers of $1$s, occurring in the $\COPT$ zone.
We allocate $v$ to the $\OSPT$ zone of $\beta$.

Then, let us consider $\beta$ to be one for simulating $y_i \in Y$ assigned $1$, 
where $1 \leq i \leq k_2$.
In each of the $\XAPT$, $\XOPT$, and $\XTPT$ zones,
we set every bit to $0$.
This means that $b$ is not for simulating the behavior of a variable in $X$.
In each of the $\YOPT$ and $\YTPT$ zones,
we set the $i$-th bit to $1$ and the other bits to $0$s.
This means that $y_i$ is assigned a Boolean value
but the other variables are not assigned. 
Note that, in more general, we do not construct the ``$\mathit{YA}$'' part
since our concern is on the counting of partial assignments 
for variables in $X_\varphi$.
In each of the $\COPT$ and $\CTPT$ zones,
for every $1 \leq j \leq k_3$,
we set the $j$-th bit to $1$
if the clause $C_j$ includes the literal $y_i$;
and the other bits to $0$s otherwise.
Let $v$ be the binary representation of the integer $2 + 2l$
such that $l$ is the numbers of $1$s, occurring in the $\COPT$ zone.
We allocate $v$ to the $\OSPT$ zone of $\beta$.
Similarly, we can consider the construction in the case where
$\beta$ is one for simulating $y_i$ assigned $0$.

\begin{figure}[htbp]
\begin{center}
\includegraphics{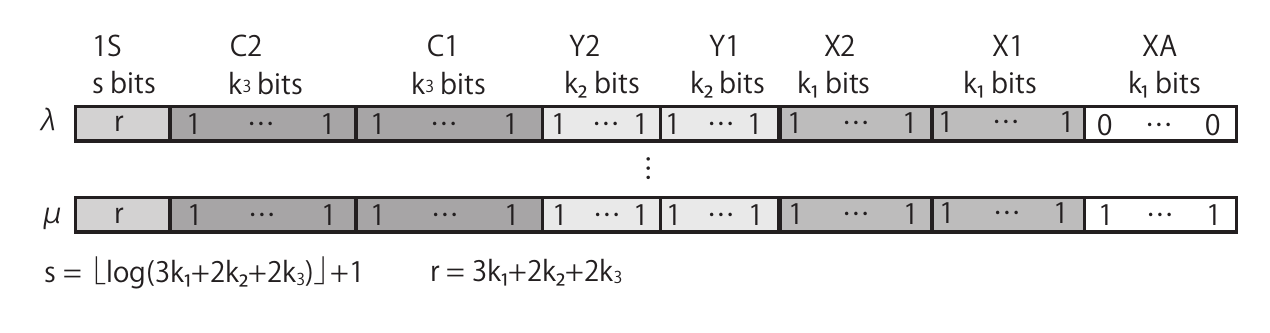}
\end{center}
\caption{Interval $[\lambda, \mu]$ that simulates assignments in $\varphi$,
where $\lambda = (r+1)2^{3k_1+2k_2+2k_3} - 2^k$
and $\mu = (r+1)2^{3k_1+2k_2+2k_3} - 1$.}
\label{fig:lambda_and_mu}
\end{figure}

As described above,
we simulate an assignment to a variable with an integer.
Moreover, for every variable, we construct two integers 
that correspond to assigned values.
Let $A$ be the set of $(k_1+k_2)$ integers such that
for every variable in $X_\varphi \cup Y_\varphi$, 
exactly one of the two constructed integers is in $A$.
Let $\alpha$ be the binary representation of $\sum_{a \in A} a$.
Then, we can simulate an assignment for $\varphi$ with $\alpha$.
Every bit in the $\XOPT$, $\XTPT$, $\YOPT$, and $\YTPT$ zones of $\alpha$
is $1$.
The $\OSPT$ zone represent the integer $3\#_1(w_1) + 2\#_1(w_2) + 2\#_1(w_3)$ 
such that $w_1$, $w_2$, and $w_3$ are
the $\XOPT$, $\YOPT$, and $\COPT$ zones, respectively.
The $\XAPT$ zone corresponds a partial assignment for $X_\varphi$.
Moreover, if every clause contains exactly one true literal,
then every bit in the $\COPT$ and $\CTPT$ zones is $1$.
That is, for every $X$-assignment $\sigma_x$, 
if there is a $Y$-assignment $\sigma_y$ such that
exactly one true literal exists in every clause,
then the assignment $\sigma_x \sigma_y$
is simulated as an integer in the interval $[\lambda,\mu]$,
as illustrated in Figure~\ref{fig:lambda_and_mu}.

The above description suggests that
we can simulate all assignments for $\varphi$
if we correctly select exactly one of the corresponding two integers
for every variable in $X$ and $Y$.
However,
the reverse direction is more complicated.
It is more difficult to verify
whether for every variable, 
exactly one of the corresponding two integers is selected,
since, in $\NUM\IGAP$, we can repeatedly select an input integer.
Moreover, for a variable $z$,
the two integers constructed from the pair of $z$ and $0$ and the pair of $z$ and $1$
may be simultaneously selected.
Another awkward phenomenon is
carrying over in the additions of the integers that we select.
Overcoming these difficulties motivates
the introduction of the $\OSPT$ zone.
Let $a$ be the sum of some integers constructed from $\varphi$
and $\alpha$ be the binary representation of $a$.
If $a$ is in $[\lambda,\mu]$
and the $\OSPT$ zone of $\alpha$ represents
the integer $3\#_1(w_1) + 2\#_1(w_2) + 2\#_1(w_3)$ 
such that $w_1$, $w_2$, and $w_3$ are
the $\XOPT$, $\YOPT$, and $\COPT$ zones, respectively,
then we can consider that 
no carry occurs in the additions for obtaining $a$. 

In Table~\ref{tab:constructed_integers_from_varphi_one},
we can check that for every pair $(\sigma_x, \sigma_y)$ of an $X$-assignment 
$\sigma_x \in \{0,1\}^4$ and a $Y$-assignment $\sigma_y \in \{0,1\}^3$,
every clause of $\varphi_1$ includes exactly one true literal
if and only if the sum of selected integers is in the interval $[\lambda,\mu]$
in Figure~\ref{fig:lambda_and_mu}.

\subsection{Formal discussion for Theorem~\ref{thm:2}}
\label{subsec:proof_completeness_IGAP}

In this subsection, we prove Theorem~\ref{thm:2}.
As a preparation, 
we formally define the zones of the binary representation of a positive integer,
as informally described in Section~\ref{subsec:idea_reduction_to_gap_counting_problem_with_intervals}.
Let $\beta$ be the binary representation of a positive integer.
Then, we call the bit sequence from the $1$st bit to the $k_1$-th bit
the {\itshape $\XAPT$} zone of $\beta$;
the sequence from the $(k_1+1)$-th bit to the $(2k_1)$-th bit
the $\XOPT$ zone of $\beta$;
the sequence from the $(2k_1+1)$-th bit to the $(3k_1)$-th bit
the $\XTPT$ zone of $\beta$;
the sequence from the $(3k_1+1)$-th bit to the $(3k_1+k_2)$-th bit
the $\YOPT$ zone of $\beta$;
the sequence from the $(3k_1+k_2+1)$-th bit to the $(3k_1+2k_2)$-th bit
the $\YTPT$ zone of $\beta$;
the sequence from the $(3k_1+2k_2+1)$-th bit to the $(3k_1+2k_2+k_3)$-th bit
the $\COPT$ zone of $\beta$;
the sequence from the $(3k_1+2k_2+k_3+1)$-th bit to the $(3k_1+2k_2+2k_3)$-th bit
the $\CTPT$ zone of $\beta$;
and the sequence from the $(3k_1+2k_2+2k_3+1)$-th bit 
to the $(3k_1+2k_2+2k_3+\lfloor\log(3k_1+2k_2+2k_3)\rfloor+2)$-th bit
the $\OSPT$ zone of $\beta$.

\begin{figure}[htbp]
\begin{center}
\includegraphics{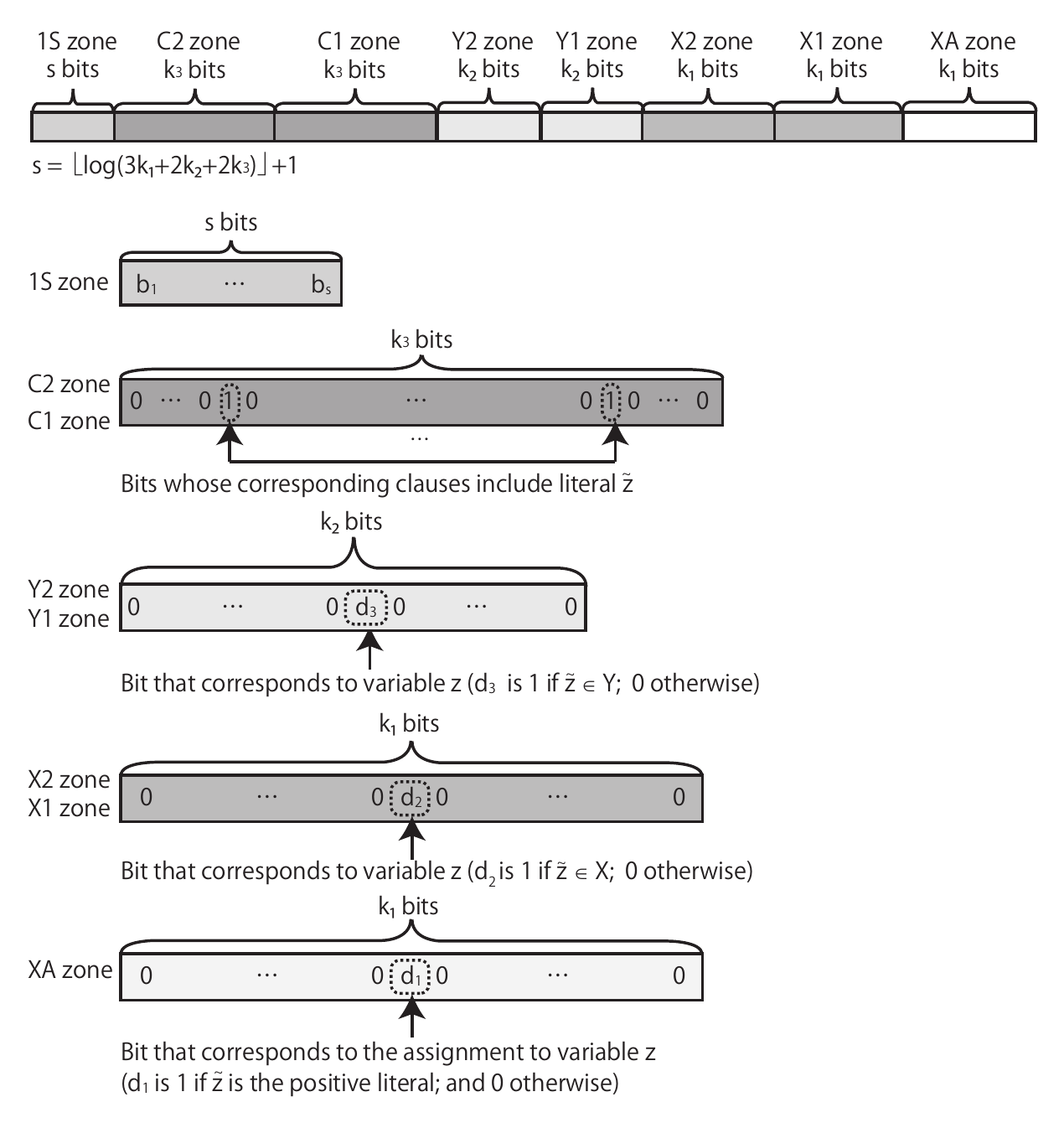}
\end{center}
\caption{Constructed integer that simulates a literal $\tilde{z}$ in a $3$-CNF formula $\varphi$.}
\label{fig:constructed_integer_for_simulating_Boolean_formula}
\end{figure}

\noindent {\itshape Proof of Theorem~\ref{thm:2}.}
Since $\NUM\NP=\NUMD{\CONP}$ \cite{TodaPhD1991English},
we may prove that $\NUM\IGAP$ is $\NUMD{\CONP}$-complete.
We first show that $\NUM\IGAP$ is in $\NUMD{\CONP}$.
Let $(A, [\kappa_0,\kappa_1])$ be an instance of $\NUM\IGAP$.
Let $M^L$ be an oracle machine, where $L$ is the decision version of
the integer knapsack problem~(\cite{Papadimitriou:1982:COA:31027}, Section 15.7).
This problem was proved to be $\NP$-complete~(\cite{Papadimitriou:1982:COA:31027}, Section 15.7).
First $M^L$ guesses a gap $e$ of $\mathcal{S}(A)$ 
in the interval $[\kappa_0,\kappa_1]$.
Then, $M^L$ verifies $e$ to be in $N(A) \cap [\kappa_0,\kappa_1]$
by using oracle $L$.
Consequently, $\NUM\IGAP$ is in $\NUMD{\CONP}$.

In the rest of the proof,
we show that $\NUM\IGAP$ is $\NUMD{\CONP}$-hard under parsimonious reductions
by reducing $\NUMD{\PI{1}\OT\SAT}$ to $\NUM\IGAP$.
In particular, we show a polynomial-time computable one-to-one mapping $s$ 
as a parsimonious reduction.
We fix $\varphi$ to be a $3$-CNF formula over $X \cup Y$
and define an instance $s(\varphi)$ of $\NUM\IGAP$,
where $X \cap Y = \emptyset$. 

Let $k_1$ and $k_2$ be the integers $|X_\varphi|$ and $|Y_\varphi|$, respectively.
Let $C_1,\cdots,C_{k_3}$ be all clauses of $\varphi$.
By the assumption in Section~\ref{subsec:2_2},
$k_3$ is larger than $1$.
We suppose that $X_\varphi = \{x_1,\cdots,x_{k_1}\}$
and $Y_\varphi = \{y_1,\cdots,y_{k_2}\}$.
We will define $(H,[\lambda,\mu])$ as an instance of $\NUM\IGAP$ below,
where $H$ is a set of positive integers
and $[\lambda,\mu]$ is an interval of positive integers.
Every integer in $H$ is 
of the form in Figure~\ref{fig:constructed_integer_for_simulating_Boolean_formula}. 

We define $\DUMMY_0$ as the integer $2^{3k_1+2k_2+2k_3}$.
For every $\tilde{x}_i \in \tilde{X}_\varphi$, 
we define an integer $h(\tilde{x}_i)$ as follows. 
\begin{eqnarray*}
h(\tilde{x}_i) & = 
& \left(|\{C_j \colon \tilde{x}_i \in C_j, 1 \leq j \leq k_3 \}| 
+ 3 \right) \DUMMY_0
\\
&& + \sum_{\substack{\tilde{x}_i \in C_j\\ 1 \leq j \leq k_3}} 
\left(2^{3k_1+2k_2+k_3+j-1} + 2^{3k_1+2k_2+j-1}\right) 
\\
&& + 2^{2k_1+i-1} + 2^{k_1+i-1} + 2^{i-1} \llbracket \tilde{x}_i = x_i \rrbracket.
\end{eqnarray*}
Similarly, for every $\tilde{y}_i \in \tilde{Y}_\varphi$, 
we define an integer $h(\tilde{y}_i)$
as follows.
\begin{eqnarray*}
h(\tilde{y}_i) & = 
& \left(|\{C_j \colon \tilde{y}_i \in C_j, 1 \leq j \leq k_3 \}|
+ 2 \right) \DUMMY_0 \\
&& + \sum_{\substack{\tilde{y_i} \in C_j\\ 1 \leq j \leq k_3}} 
\left(2^{3k_1+2k_2+k_3+j-1} + 2^{3k_1+2k_2+j-1}\right) 
\\
&& + 2^{3k_1+k_2+i-1} + 2^{3k_1+i-1}.
\end{eqnarray*}
We define $H$ as $\{h(\tilde{z}) \colon \tilde{z} \in \tilde{X}_\varphi \cup \tilde{Y}_\varphi\}$.
We define positive integers $\lambda$ and $\mu$ as follows.
\begin{eqnarray*}
&&
\lambda = 
(3k_1+2k_2+2k_3) \DUMMY_0
+ \sum_{i=k_1}^{3k_1+2k_2+2k_3-1} 2^i,\\
&&
\mu = (3k_1+2k_2+2k_3+1) \DUMMY_0 - 1.
\end{eqnarray*}
We define $s(\varphi)$ as $(H \cup \{\DUMMY_0\}, [\lambda, \mu])$.

Then, we check the time complexity of the reduction.
\ref{alg:1} shows the whole of the reduction. 
In this algorithm, 
$v$ denotes $h(\tilde{z})$ for every 
$\tilde{z} \in \tilde{X}_\varphi \cup \tilde{Y}_\varphi$ 
where $\tilde{z}$ denotes 
$x_i$ if $Z = X_\varphi$ and $p = 1$;
$\lnot x_i$ if $Z = X_\varphi$ and $p = 0$;
$y_i$ if $Z = Y_\varphi$ and $p = 1$;
and $\lnot y_i$ otherwise.
By \ref{alg:1}, 
the reduction can compute $(H \cup \{\DUMMY_0\}, [\lambda, \mu])$
in polynomial time.

\begin{algorithm}
\caption{Reduction from $\NUMD\PI{1}\OT\SAT$ to $\IGAP$}
\label{alg:1}
\begin{algorithmic}
 \STATE {\bfseries Input.} A $3$-CNF formula 
 $\varphi = (C_1 \land \cdots \land C_{k_3})$ over $X \cup Y$, 
 where $|X_\varphi| = k_1$ and $|Y_\varphi| = k_2$.
 \STATE {\bfseries Output.} $(H \cup \{\DUMMY_0\}, [\lambda, \mu])$.
  \STATE $\DUMMY_0 \leftarrow 2^{3k_1+2k_2+2k_3}$
  \STATE $\lambda \leftarrow (3k_1+2k_2+2k_3) \DUMMY_0 + \sum_{i=k_1}^{3k_1+2k_2+2k_3-1} 2^i$
  \STATE $\mu \leftarrow (3k_1+2k_2+2k_3+1) \DUMMY_0 - 1$
  \STATE $H \leftarrow \emptyset$
  \FOR{each $\mathit{Z} \in \{X_\varphi,Y_\varphi\}$}
   \FOR{$i = 1$ to $|Z|$} 
    \FOR{$p \in \{0,1\}$} 
     \STATE $c \leftarrow 2$
     \IF{$Z=X_\varphi$}
      \STATE $v \leftarrow 2^{2k_1+i-1} + 2^{k_1+i-1}$
      \IF{$p=1$}
       \STATE $v \leftarrow v + 2^{i-1}$
       \STATE $c \leftarrow c + 1$
       \STATE $\tilde{z} \leftarrow x_i$
      \ELSE
       \STATE $\tilde{z} \leftarrow \lnot x_i$
      \ENDIF
     \ELSE
      \STATE $v \leftarrow 2^{3k_1+k_2+i-1} + 2^{3k_1+i-1}$
      \IF{$p = 1$}
       \STATE $\tilde{z} \leftarrow y_i$
      \ELSE
       \STATE $\tilde{z} \leftarrow \lnot y_i$
      \ENDIF
     \ENDIF
     \FOR{$j = 1$ to $k_3$}
      \IF{$\tilde{z} \in C_j$}
       \STATE $v \leftarrow v + 2^{3k_1+2k_2+k_3+j-1} + 2^{3k_1+2k_2+j-1}$
       \STATE $c \leftarrow c + 1$
      \ENDIF
     \ENDFOR
     \STATE $v \leftarrow v + c \times \DUMMY_0$
     \STATE $H \leftarrow H \cup \{v\}$
    \ENDFOR 
   \ENDFOR 
  \ENDFOR
  \RETURN $(H, [\lambda,\mu])$
\end{algorithmic} 
\end{algorithm} 

\begin{claim}
\label{cla:10}
$s(\varphi)$ can be constructed in polynomial time on the size of $\varphi$.
\end{claim}
{\itshape Proof of Claim~\ref{cla:10}.}
It suffices to check $\lambda$, $\mu$, and all integers in $H$ to be computable
in time polynomial in $k_1$, $k_2$, and $k_3$.
Let $\tilde{z}$ be a literal in $\tilde{X}_\varphi \cup \tilde{Y}_\varphi$.
The integer $h(\tilde{z})$ consists of
at most $3k_1+2k_2+2k_3 + \lfloor \log(3k_1+2k_2+2k_3)\rfloor + 1$ bits.
We can construct
each of the $\XAPT$, $\XOPT$, $\XTPT$, $\YOPT$, and $\YTPT$ zones of $h(\tilde{z})$
in time linear in $k_1$, $k_2$ and $k_3$.
For every clause $C$ in $\varphi$,
we can check whether $\tilde{z}$ is in $C$ in constant polynomial. 
Thus, we construct 
the $\COPT$ and $\CTPT$ zones of $h(\tilde{z})$
in time polynomial in $k_1$, $k_2$, and $k_3$.
After all the zones except the $\OSPT$ zone are constructed,
we can count $1$s in all the zones except the $\OSPT$ zone
in time linear in $k_1$, $k_2$, and $k_3$.
Thus, we can calculate the $\OSPT$ zone 
in time polynomial in $k_1$, $k_2$, and $k_3$.
It follows that $h(\tilde{z})$ is computed in time polynomial in $k_1$, $k_2$, and $k_3$.
The integer $h(\tilde{z})$ is calculated in time polynomial in $k_1$, $k_2$, and $k_3$.
The set $H$ has $2(k_1+k_2)$ elements
since an integer is defined as an element
for every literal in $\tilde{X}_\varphi \cup \tilde{Y}_\varphi$.
By the similar method in the above discussion,
the integers $\lambda$ and $\mu$ 
can be computed in time polynomial in $k_1$, $k_2$, and $k_3$.
Consequently, the claim holds. 
\QED (Claim~\ref{cla:10})

We will check the validity of the above reduction.
We show that there is exactly one true literal in every clause in $\varphi$ 
for a given assignment $\sigma$
if and only if
there is a set $S \subseteq H \cup \{\DUMMY_0\}$ such that
$\sum_{\tau \in S} \tau$ is in $[\lambda, \mu]$.
The {\itshape ``only if''} part
is straightforward by definition.
We show the {\itshape ``if''} part here.
In the rest of the proof for Theorem~\ref{thm:2},
we fix $\alpha_1,\cdots,\alpha_\iota$ to be positive integers in $H$,
and $\alpha$ to be the sum of $\alpha_1,\cdots,\alpha_\iota$,
where $\iota \geq 1$.
For every $\tilde{z} \in \tilde{X}_\varphi \cup \tilde{Y}_\varphi$,
we denote the number $|\{i \colon \alpha_i = h(\tilde{z}), 1 \leq i \leq \iota\}|$ 
by $c(\tilde{z})$.

\begin{claim}
\label{cla:1}
For every $z \in X_\varphi \cup Y_\varphi$,
exactly one of the following conditions holds.
\begin{enumerate}
\item $c(z) = 0$ and $c(\lnot z) = 1$.
\item $c(z) = 1$ and $c(\lnot z) = 0$.
\end{enumerate}
\end{claim}

To prove Claim~\ref{cla:1}, it show a more general statement
in Claim~\ref{cla:2}. 
As preparation, we define some notions and notations for any nonnegative integer $e$.
For every $1 \leq i \leq 3$,
let $I_{i,e}$ be a set of integers in $[1,k_i]$.
Let $I_{0,e}$ be a subset of $I_{1,e}$.
We say that $e$ is {\itshape consistent} with $\varphi$ 
if $e$ can be written as
\begin{eqnarray*}
&& \left(3|I_{1,e}| + 2|I_{2,e}| + 2|I_{3,e}|\right) \DUMMY_0
+ \sum_{i \in I_{3,e}} (2^{k_3} + 1) 2^{3k_1+2k_2+i-1}
\\
&& 
+ \sum_{i \in I_{2,e}} (2^{k_2} + 1) 2^{3k_1+i-1}
+ \sum_{i \in I_{1,e}} (2^{k_1} + 1) 2^{k_1+i-1}
+ \sum_{i \in I_{0,e}} 2^{i-1}.
\end{eqnarray*}
Note that if $e$ is in $[\lambda,\varphi]$,
then $e$ is consistent with $\varphi$.
For every $1 \leq i \leq \iota$,
we denote the sum $\sum_{i = 1}^ i e_i$ by $\xi_i$.
We may assume the addition of $e_1, \cdots, e_\iota$
to do in the left associative manner; i.e.,
$((e_1+e_2)+ \cdots + e_{\iota-1})+e_\iota$,
without loss of generality.
Claim~\ref{cla:4} is essential 
although it can be immediately obtained from definitions.

\begin{claim}
 \label{cla:4}
$\xi_i$ is consistent with $\varphi$
for every $1 \leq i \leq \iota$
if and only if
no carry occurs at 
the $\XAPT$, $\XOPT$, $\XTPT$, $\YOPT$, $\YTPT$, $\COPT$, and $\CTPT$ zones
in the addition of $\xi_{i-1}$ and $\alpha_i$
for every $2 \leq i \leq \iota$.
\end{claim}
\noindent {\itshape Proof of Claim~\ref{cla:4}.}
The {\itshape ``if ''} part is trivial 
by the definitions of integers in $H$.
Thus, we prove the {\itshape ``only if''} part below.
The proof is by induction on $\iota \geq 1$.
Suppose that $\xi_i$ is consistent with $\varphi$
for every $1 \leq i \leq \iota$.
By induction hypothesis,
no carry occurs at 
the $\XAPT$, $\XOPT$, $\XTPT$, $\YOPT$, $\YTPT$, $\COPT$, and $\CTPT$ zones
in the addition $\xi_{i} + \alpha_{i-1}$
for every $2 \leq i \leq \iota-1$.
By definition,
every element of $H$ is consistent with $\varphi$.
Thus, $\alpha_i$ is consistent with $\varphi$.
To show that
no carry occurs at 
the $\XAPT$, $\XOPT$, $\XTPT$, $\YOPT$, $\YTPT$, $\COPT$, and $\CTPT$ zones
in the addition $\xi_{i-1} + \alpha_i$,
we separately describe the cases depending on
whether a carry occurs at one of 
the $\XOPT$, $\XTPT$, $\YOPT$, $\YTPT$, $\COPT$, and $\CTPT$ zones
in that addition.
First, assume that a carry occurs at one of the 
the $\XOPT$, $\XTPT$, $\YOPT$, $\YTPT$, $\COPT$, and $\CTPT$ zones
in that addition.
Then, 
\begin{center}
$\#_1(\xi_\iota\mod \DUMMY_0) 
< \#_1(\xi_{\iota-1} \mod \DUMMY_0) + \#_1(\alpha_\iota \mod \DUMMY_0)$.
\end{center}
Since  $\xi_\iota$ is consistent with $\varphi$,
\begin{center}
$\#_1(\xi_\iota \mod \DUMMY_0)
= \#_1(\xi_{\iota-1} \mod \DUMMY_0) + \#_1(\alpha_\iota \mod \DUMMY_0)$.
\end{center}
However, in the addition $\xi_{\iota-1} + \alpha_\iota$,
if a carry occurs at the $\XAPT$ zone,
then at least two carries occur 
at the $\XOPT$ and $\XTPT$ zones
since $I_{0,\xi_{i-1}} \subseteq I_{1,\xi_{i-1}}$
and $I_{0,\alpha_i} \subseteq I_{1,\alpha_i}$
by their consistencies.
It contradicts to the assumption.
Hence, no carry occurs at the $\XAPT$ zone.
Next, assume that no carry occurs at 
the $\XOPT$, $\XTPT$, $\YOPT$, $\YTPT$, $\COPT$, and $\CTPT$ zones
in the addition $\xi_{\iota-1} + \alpha_\iota$.
Then, by the definitions of the $\XAPT$, $\XOPT$, and $\XTPT$ zones,
no carry occurs at the $\XAPT$ zone.
Consequently, 
no carry occurs at 
the $\XAPT$, $\XOPT$, $\XTPT$, $\YOPT$, $\YTPT$, $\COPT$, and $\CTPT$ zones
in the addition $\xi_{i-1} + \alpha_i$
for every $2 \leq i \leq l$.
The proof is complete.
\QED (Claim~\ref{cla:4})

The following claim follows from Claim~\ref{cla:4}.

\begin{claim}
\label{cla:2}
If $\alpha$ is consistent with $\varphi$,
then the following statements hold.
\begin{enumerate}
\item For every $x_j \in X_\varphi$ with $1 \leq j \leq k_1$,
\begin{itemize}
\item if $j \in I_{0,\alpha} \cap I_{1,\alpha}$, then
      $c(x_j) = 1$ and $c(\lnot x_j) = 0$,
\item if $j \in I_{1,\alpha} \backslash I_{0,\alpha}$, then
      $c(x_j) = 0$ and $c(\lnot x_j) = 1$,
\item otherwise $c(x_j) = c(\lnot x_j) = 0$. 
\end{itemize}

\item For every $y_j \in Y_\varphi$ with $1 \leq j \leq k_2$,
\begin{itemize}
\item if $j \in I_{2,\alpha}$, then
      $c(\tilde{y}_j) = 1$ and $c(\tilde{y}_j^c) = 0$,
\item otherwise $c(y_j) = c(\lnot y_j) = 0$. 
\end{itemize}
\end{enumerate}
\end{claim}

Consequently, the theorem holds.
\QED (Theorem~\ref{thm:3})

\section{Complexity of $\NUM \TGAP$}
\label{sec:proof_bounded_gaps_parsimonious_reducibility}

In Section~\ref{sec:proof_parsimonious_reducibility},
we proved 
$\NUMD{\PI{1}\OT\SAT}$ to be parsimonious reducible to $\NUM\IGAP$.
In this section, we prove the following theorem
by extending the reduction in the proof for Theorem~\ref{thm:2} 
to one from $\NUMD{\PI{1}\OT\SAT}$ to $\NUM\TGAP$.
\begin{theorem}
\label{thm:3} 
$\NUM\TGAP$ is $\NUM\NP$-complete under parsimonious reductions.
\end{theorem}

\subsection{Ideas of the proof}
\label{subsec:observation_TGAP}

In this subsection,
let $\varphi$ be fixed to a $3$-CNF formula over $X \cup Y$,
as in the proof for Theorem~\ref{thm:2},
where $X$ and $Y$ are disjoint.
Similarly, let $\lambda, \mu, H, h$, and $\DUMMY_0$ 
be defined as in the proof for Theorem~\ref{thm:2}.
For extending the reduction in the proof for Theorem~\ref{thm:2} 
to one from $\NUMD{\PI{1}\OT\SAT}$ to $\NUM\TGAP$,
we specify all gaps of $\mathcal{S}(H \cup \{\DUMMY_0\})$
in $[\lambda, \mu]$
by using another set $s_0(\varphi)$ that contains $H$ as a subset.
In more detail, 
we construct the mapping $s_0$ so that
$N(H \cup \{\DUMMY_0\}) \cap [\lambda,\mu] = N(s_0(\varphi)) \cap [\lambda,\infty]$.

\subsubsection{Observation}
\label{subsubsec:511}

\begin{figure}[htbp]
\begin{center}
\includegraphics{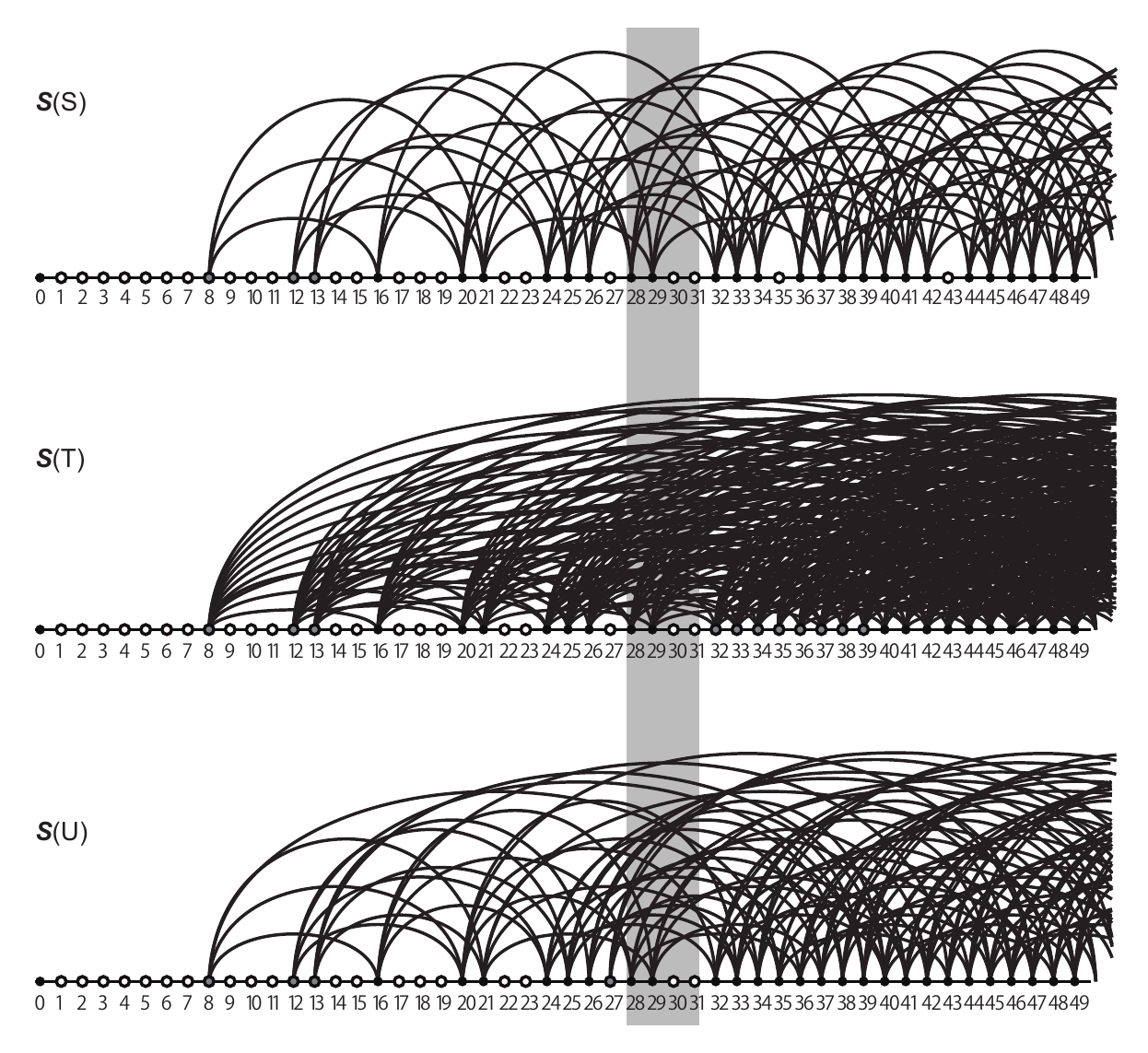}
\end{center}
\caption{Numerical semigroups generated by $S = \{8,12,13\}$, $T = \{8,12,13,32,33,34,35,36,37,38,39\}$, and $U = \{8,12,13,27\}$.}
\label{fig:observation_for_TGAP}
\end{figure}

First, we describe how to specify every integer 
in $N(H \cup \{\DUMMY_0\}) \cap [\lambda,\mu]$ 
as a gap of another numerical semigroup, 
which is larger than a bound.
Let us first consider a toy example of $\NUM\IGAP$
instead of an input constructed 
from a $3$-CNF formula.
Let $S$ be the set $\{8,12,13\}$.
We observe an example illustrated in Figure~\ref{fig:observation_for_TGAP}.
In Figure~\ref{fig:observation_for_TGAP}, 
all the circles on the three lines mean integers in $[0,49]$.
On every line, the grayed circles mean positive integers given as inputs.
The black circles mean positive integers, each of which can be written
as a nonnegative integer combination of the input positive integers
but is not one of the input positive integers themselves.
The white circles mean other integers in $[0,49]$;
i.e., integers that cannot be written
as nonnegative integer combinations of the corresponding input integers.
The circles on the $1$st line illustrate a part of $\mathcal{S}(S)$.
That is, we can observe an instance $(S, [28,31])$ of $\NUM\TGAP$
on the $1$st line.

Let us next specify two gaps $30$ and $31$ of $\mathcal{S}(S)$ in $[28,31]$
with another numerical semigroup and only the lower bound $28$
but without the upper bound $31$. 
For this purpose, we may find a numerical semigroup $\mathcal{B}$ such that 
$(\mathbb{N} \backslash \mathcal{B}) \cap [28,\infty] = N(S) \cap [28,31]$.
A simple but naive approach to specify $\mathcal{B}$ is
by using $T = S \cup [32,39]$ as its generators.
The $2$nd line of Figure~\ref{fig:observation_for_TGAP}
illustrates $\mathcal{S}(T)$.
By definition, $T$ contains successive integers from $32$ to $39$.
The number of these successive integers is not smaller than
the smallest element of $T$.
This fact guarantees that $N(T)$ does not include any integer greater than $31$.
Unfortunately, the size of $T$ is of exponential order in the size of $S$
since we assume the size of an integer to be its bit length.
This defect is caused by the cardinality of $T$.

For making the number of generators of $\mathcal{B}$ smaller,
it suffices to specify every integer between $32$ and $39$
as a nonnegative integer combination of a smaller set of positive integers.
More generally,
it is not so difficult 
since we can specify every integer in an interval $[\kappa,\kappa+m-1]$
of positive integers
as a nonnegative integer combination of $l$ positive integers for some $l = O(\log m)$.
However,
we have to be careful to preserve all gaps of $\mathcal{S}(S)$ between $28$ and $31$.
If $U$ is the set $S \cup \{27\}$, then
$U$ satisfies the above conditions.
As illustrated on the $3$rd line in Figure~\ref{fig:observation_for_TGAP}, 
all integers in $N(S)$ greater than $31$
are included by $N(U)$, which are $35$ and $43$,
and $N(S)$ does not contain any gap of $\mathcal{S}(S)$ in $[28,31]$, which are $30$ and $31$.

\subsubsection{An example of our reduction}
\label{subsubsec:instance_reduction_TGAP}

Next, we illustrate the reduction that we will formulate in Section~\ref{subsec:proof_TGAP}
by using an example.
Let $\varphi_2$ be a $3$-CNF formula $C_1 \land C_2$, where
$C_1 = \lnot x_1 \lor x_2 \lor y_1$ and $C_2 = x_1 \lor x_2 \lor y_2$.
Table~\ref{tab:3} shows clauses, each of which contains exactly one true clause
for every assignment for $\varphi_2$.

\begin{table}[htbp]
\begin{center}
\caption{Clauses in $\varphi_2$, each of which contains exactly one true literal,
where each cell contains indices of clauses for an assignment.}
\label{tab:3}
\begin{tabular}{ccccc}
\hline
 $\sigma_x$ & \multicolumn{4}{c}{$\sigma_y$} \\
 \cline{2-5}
 & $00$ & $10$ & $01$ & $11$ \\
 \hline
\rowcolor{MyGray} $00$ & $1$ &  --- & $1,2$ &  $2$  \\
\rowcolor{MyGray} $10$ & $2$ &  $1,2$ &  --- & $1$ \\
 $01$ & $2$ &  $2$ &  --- & --- \\
 $11$ & $1$ &  --- & $1$ &  --- \\
 \hline
 \end{tabular}
\end{center}
\end{table}

\begin{table}[htbp]
\caption{Integers constructed from $\varphi_2$ by the method in the proof for Theorem~\ref{thm:2}.}
\label{tab:A_for_varphi_two}
\begin{center}
\begin{tabular}{cc|cccccccc}
\hline
\multicolumn{2}{c|}{Arguments of $s$} 
& \multicolumn{8}{c}{Constructed integers} \\
\hline
Variables & Values
& $\OSPT$ & $\CTPT$ & $\COPT$ 
& $\YTPT$ & $\YOPT$ & $\XTPT$ & $\XOPT$ & $\XAPT$
\\
\hline
$x_1$ & $1$ & $5$ & $10$ & $10$ & $00$ & $00$ & $01$ & $01$ & $01$\\
$x_1$ & $0$ & $5$ & $01$ & $01$ & $00$ & $00$ & $01$ & $01$ & $00$\\
$x_2$ & $1$ & $7$ & $11$ & $11$ & $00$ & $00$ & $10$ & $10$ & $10$\\
$x_2$ & $0$ & $3$ & $00$ & $00$ & $00$ & $00$ & $10$ & $10$ & $00$\\
$y_1$ & $1$ & $4$ & $01$ & $01$ & $01$ & $01$ & $00$ & $00$ & $00$\\
$y_1$ & $0$ & $2$ & $00$ & $00$ & $01$ & $01$ & $00$ & $00$ & $00$\\
$y_2$ & $1$ & $4$ & $10$ & $10$ & $10$ & $10$ & $00$ & $00$ & $00$\\
$y_2$ & $0$ & $2$ & $00$ & $00$ & $10$ & $10$ & $00$ & $00$ & $00$\\
\hline
\end{tabular}
\end{center}
\end{table}

\begin{table}[htbp]
\caption{Endpoints of the interval constructed from $\varphi_2$ by the method in the proof for Theorem~\ref{thm:2}.}
\label{tab:lambda_mu_for_varphi_two}
\begin{center}
\begin{tabular}{c|cccccccc}
\hline
\multirow{2}{*}{Endpoints} & \multicolumn{7}{c}{Constructed integers} &\\
\cline{2-9}
& $\OSPT$ & $\CTPT$ & $\COPT$ 
& $\YTPT$ & $\YOPT$ & $\XTPT$ & $\XOPT$ & $\XAPT$
\\
\hline
$\lambda_{\varphi_2}$ & $14$ & $11$ & $11$ & $11$ & $11$ & $11$ & $11$ & $00$ \\
$\mu_{\varphi_2}$ & $14$ & $11$ & $11$ & $11$ & $11$ & $11$ & $11$ & $11$ \\
\hline
\end{tabular}
\end{center}
\end{table}

Tables \ref{tab:A_for_varphi_two} and \ref{tab:lambda_mu_for_varphi_two}
show integers constructed from $\varphi_2$
by using the reduction
in the proof for Theorem~\ref{thm:2}.
Tables \ref{tab:D_for_varphi_two} and \ref{tab:D_plus_xk1_for_varphi_two}
show integers that we will construct. 
Moreover, we define $\DUMMY_{0,\varphi_2}$ as 
the integer $2^{14}$.
This is defined from $\varphi_2$
by the same method for $\DUMMY_0$
in the reduction in the proof for Theorem~\ref{thm:2}.

The numerical semigroup generated by the integers
in Tables \ref{tab:D_for_varphi_two} 
or \ref{tab:D_plus_xk1_for_varphi_two} satisfies the conditions that we described in \ref{subsubsec:511}.
For every integer $e \geq \lambda_{\varphi_2}$,
if $e$ is greater than $\mu_{\varphi_2}$,
then $e$ can be represented as a nonnegative integer combination 
of $\DUMMY_{0,\varphi_2}$
and integers in Tables \ref{tab:D_for_varphi_two} or \ref{tab:D_plus_xk1_for_varphi_two};
and otherwise, $e$ cannot be represented by the same way.
Intuitively speaking,
every integer in
Tables \ref{tab:D_for_varphi_two} or \ref{tab:D_plus_xk1_for_varphi_two}
is a kind of dummy.
In Tables \ref{tab:D_for_varphi_two} and \ref{tab:D_plus_xk1_for_varphi_two},
every row is tagged with a variable or a clause in the $1$st column,
although any integer
does not have a role for simulating a $3$-CNF formula.
If we select integers, one each for every variable and clause
in Tables \ref{tab:D_for_varphi_two} and \ref{tab:D_plus_xk1_for_varphi_two},
then we can represent an integer greater than $\mu_{\varphi_2}$
as their sum, and vice versa.

We describe the details of integers in 
Tables \ref{tab:D_for_varphi_two} or \ref{tab:D_plus_xk1_for_varphi_two}.
Let $\beta$ be the binary representation of an integer between the $1$st and $12$th rows
in Table~\ref{tab:D_for_varphi_two}.
$\beta$ is tagged with a variable in $X_{\varphi_2}$. 
Let $l$ be $1$ if $\beta$ is tagged with $x_1$;
and $2$ otherwise.
In each of the $\XAPT$, $\XOPT$, and $\XTPT$ zones of $\beta$,
$1$ occurs at most once.
Conversely,
in at least one of the $\XAPT$, $\XOPT$, and $\XTPT$ zones,
$1$ occurs.
In each of the $\XAPT$ and $\XTPT$ zones,
if $1$ occurs, then its position is the $l$-th bit of the zone.
In the $\XOPT$ zone,
if $1$ occurs, then its position is the $((l+1) \mod 2)$-th bit of the zone,
where the modulus $2$ is from the cardinality of $X_{\varphi_2}$.
As described above, every integer in
Tables \ref{tab:D_for_varphi_two} or \ref{tab:D_plus_xk1_for_varphi_two}
is for representing all integers larger than $\mu_{\varphi_2}$
but not for simulating a $3$-CNF formula.
Thus, for avoiding improperly simulating a $3$-CNF formula,
the position of the bit of $1$ is shifted in the $\XOPT$ zone.

Similarly,
in Tables \ref{tab:D_for_varphi_two} or \ref{tab:D_plus_xk1_for_varphi_two},
every integer between the $13$rd and $18$th rows
is tagged with a variable in $Y_{\varphi_2}$.
Moreover, every integer between the $19$th and $24$th rows
is tagged with a clause in $\varphi_2$.

For every $z \in X_{\varphi_2} \cup Y_{\varphi_2} \cup \mathcal{C}_\varphi$,
let $\beta_z$ be a binary representation in Table~\ref{tab:D_for_varphi_two}, 
which is tagged with $z$.
In $\beta_z$,
the integer represented by its $\OSPT$ zone is equal to $c$, 
where $c$ is the maximum number of occurrences of $1$s
in its remaining zones among all integers tagged with $z$ 
in Table~\ref{tab:D_for_varphi_two}. 
On the other hand,
in Table~\ref{tab:D_plus_xk1_for_varphi_two},
the integer represented in its $\OSPT$ zone is equal to $c+1$,
where $c$ is the maximum number of occurrences of $1$s in its remaining zones. 
The difference of the $\OSPT$ zones in integers 
in Tables \ref{tab:D_for_varphi_two} and \ref{tab:D_plus_xk1_for_varphi_two}
is for representing every integer greater than $\mu_{\varphi_2}$
as a sum of $\DUMMY_0$ and integers 
that we select one each for every variable and clause in $\varphi_2$.

\begin{table}[htbp]
\caption{Integers constructed from $\varphi_2$, which are introduced for the reduction to $\TGAP$.}
\label{tab:D_for_varphi_two}
\begin{center}
\begin{tabular}{cc|cccccccc}
\hline
Variables & \multirow{2}{*}{Values}
& \multicolumn{7}{c}{Constructed integers} & \\
\cline{3-10}
or clauses & & $\OSPT$ & $\CTPT$ & $\COPT$ 
& $\YTPT$ & $\YOPT$ & $\XTPT$ & $\XOPT$ & $\XAPT$
\\
\hline
\multirow{7}{*}{$x_1$} & $001$ & $3$ & $00$ & $00$ & $00$ & $00$ & $00$ & $00$ & $01$  \\
 & $010$ & $3$ & $00$ & $00$ & $00$ & $00$ & $00$ & $10$ & $00$  \\
 & $011$ & $3$ & $00$ & $00$ & $00$ & $00$ & $00$ & $10$ & $01$  \\
 & $100$ & $3$ & $00$ & $00$ & $00$ & $00$ & $01$ & $00$ & $00$  \\
 & $101$ & $3$ & $00$ & $00$ & $00$ & $00$ & $01$ & $00$ & $01$  \\
 & $110$ & $3$ & $00$ & $00$ & $00$ & $00$ & $01$ & $10$ & $00$  \\
 & $111$ & $3$ & $00$ & $00$ & $00$ & $00$ & $01$ & $10$ & $01$  \\
\hline
\multirow{5}{*}{$x_2$} & $001$ & $3$ & $00$ & $00$ & $00$ & $00$ & $00$ & $00$ & $10$   \\
 & $010$ & $3$ & $00$ & $00$ & $00$ & $00$ & $00$ & $01$ & $00$   \\
 & $011$ & $3$ & $00$ & $00$ & $00$ & $00$ & $00$ & $01$ & $10$   \\
 & $100$ & $3$ & $00$ & $00$ & $00$ & $00$ & $10$ & $00$ & $00$  \\
 & $101$ & $3$ & $00$ & $00$ & $00$ & $00$ & $10$ & $00$ & $10$  \\
\hline
\multirow{3}{*}{$y_1$} & $01$ & $2$ & $00$ & $00$ & $00$ & $10$ & $00$ & $00$ & $00$  \\
 & $10$ & $2$ & $00$ & $00$ & $01$ & $00$ & $00$ & $00$ & $00$  \\
 & $11$ & $2$ & $00$ & $00$ & $01$ & $10$ & $00$ & $00$ & $00$  \\
\hline
\multirow{3}{*}{$y_2$} & $01$ & $2$ & $00$ & $00$ & $00$ & $01$ & $00$ & $00$ & $00$  \\
 & $10$ & $2$ & $00$ & $00$ & $10$ & $00$ & $00$ & $00$ & $00$  \\
 & $11$ & $2$ & $00$ & $00$ & $10$ & $01$ & $00$ & $00$ & $00$  \\
\hline
\multirow{3}{*}{$c_1$} & $01$ & $2$ & $00$ & $10$ & $00$ & $00$ & $00$ & $00$ & $00$  \\
 & $10$ & $2$ & $01$ & $00$ & $00$ & $00$ & $00$ & $00$ & $00$  \\
 & $11$ & $2$ & $01$ & $10$ & $00$ & $00$ & $00$ & $00$ & $00$  \\
\hline
\multirow{3}{*}{$c_2$} & $01$ & $2$ & $00$ & $01$ & $00$ & $00$ & $00$ & $00$ & $00$  \\
 & $10$ & $2$ & $10$ & $00$ & $00$ & $00$ & $00$ & $00$ & $00$  \\
 & $11$ & $2$ & $10$ & $01$ & $00$ & $00$ & $00$ & $00$ & $00$  \\
\hline
\end{tabular}
\end{center}
\end{table}

\begin{table}[htbp]
\caption{Integers constructed from $\varphi_2$, which are introduced for the reduction to $\TGAP$}
\label{tab:D_plus_xk1_for_varphi_two}
\begin{center}
\begin{tabular}{cc|cccccccc}
\hline
Variables & \multirow{2}{*}{Values}
& \multicolumn{7}{c}{Constructed integers} & \\
\cline{3-10}
or clauses & & $\OSPT$ & $\CTPT$ & $\COPT$ 
& $\YTPT$ & $\YOPT$ & $\XTPT$ & $\XOPT$ & $\XAPT$
\\
\hline
\multirow{2}{*}{$x_2$} & $110$ & $4$ & $00$ & $00$ & $00$ & $00$ & $10$ & $01$ & $00$ \\
 & $111$ & $4$ & $00$ & $00$ & $00$ & $00$ & $10$ & $01$ & $10$ \\
\hline
\end{tabular}
\end{center}
\end{table}

\subsection{Formal discussion for Theorem~\ref{thm:3}}
\label{subsec:proof_TGAP}

\noindent{\itshape Proof of Theorem~\ref{thm:3}.}
First, we prove that $\NUM\TGAP$ is in $\NUMD{\PIP{1}}$, i.e., $\NUM\NP$.
By Theorem~\ref{thm:2},
it suffices to show that $\NUM\TGAP$ is a special case of $\NUM\IGAP$ as follows.
Let $(A,\kappa)$ be an instance of $\NUM\TGAP$.
Let $a$ be the maxmum element of $A$.
By definition, all integers in $A$ are coprime.
Thus, there is the maximum element $g(A)$ in $N(A)$. 
By an upper bound $a^2$ for $g(A)$ \cite{Wilf1978},
$N(A) \cap [\kappa,\infty]$ is equal to $N(A) \cap [\kappa,a^2]$.
Thus, $\NUM\TGAP$ is an special case of $\NUM\IGAP$.
Consequently, $\NUM\TGAP$ is in $\NUM\NP$.

Next, we prove that $\NUM\TGAP$ is $\NUMD{\PIP{1}}$-hard under parsimonious reductions.
For this purpose, 
we reduce $\NUMD{\PI{1}\OT\SAT}$ to $\NUM\TGAP$.
Let $X$, $Y$, $\varphi$, $k_1$, $k_2$, and $k_3$ 
be given as in the proof for Theorem~\ref{thm:2}.
$\varphi$ is a $3$-CNF formula over $X \cup Y$.
$k_1$, $k_2$, and $k_3$ are $|X_\varphi|$, $|Y_\varphi|$, and $|\mathcal{C}_\varphi|$,
respectively.
Let $h, H, [\lambda,\mu]$ and $\DUMMY_0$ be defined
as in the proof for Theorem~\ref{thm:2}.
$h$ is a polynomial-time computable function from $3$-CNF-formulae to positive integers.
$H$ is a set of positive integers.
$[\lambda, \mu]$ is an interval of positive integers.
and $\DUMMY_0$ is a positive integer.
By the proof for Theorem~\ref{thm:2},
we can construct $h, H, [\lambda,\mu]$ and $\DUMMY_0$
in polynomial time.

\begin{figure}[htbp]
\begin{center}
\includegraphics{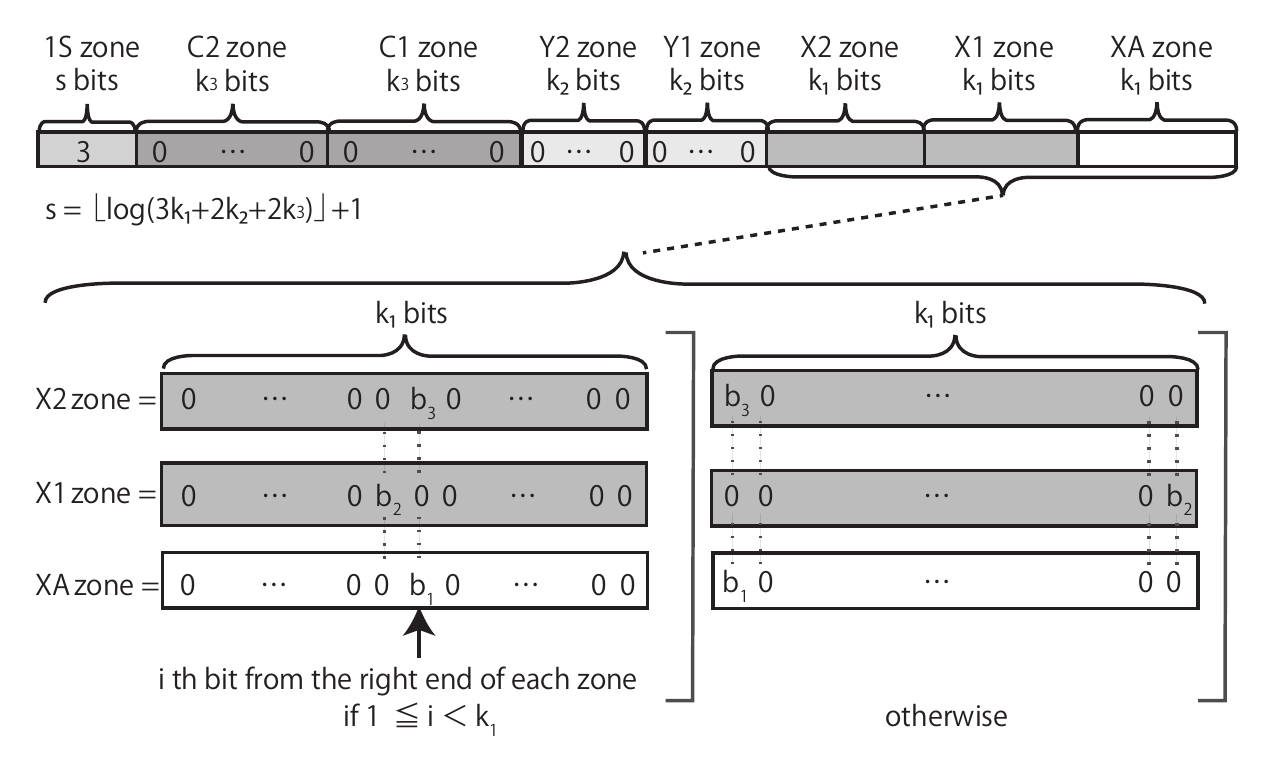}
\end{center}
\caption{Integer $d(x_i, b_2 b_1 b_0)$ constructed in the proof for Theorem~\ref{thm:3}.}
\label{fig:constructed_integer_for_an adjustment_on_a_variable_in_X}
\end{figure}

\begin{figure}[htbp]
\begin{center}
\includegraphics{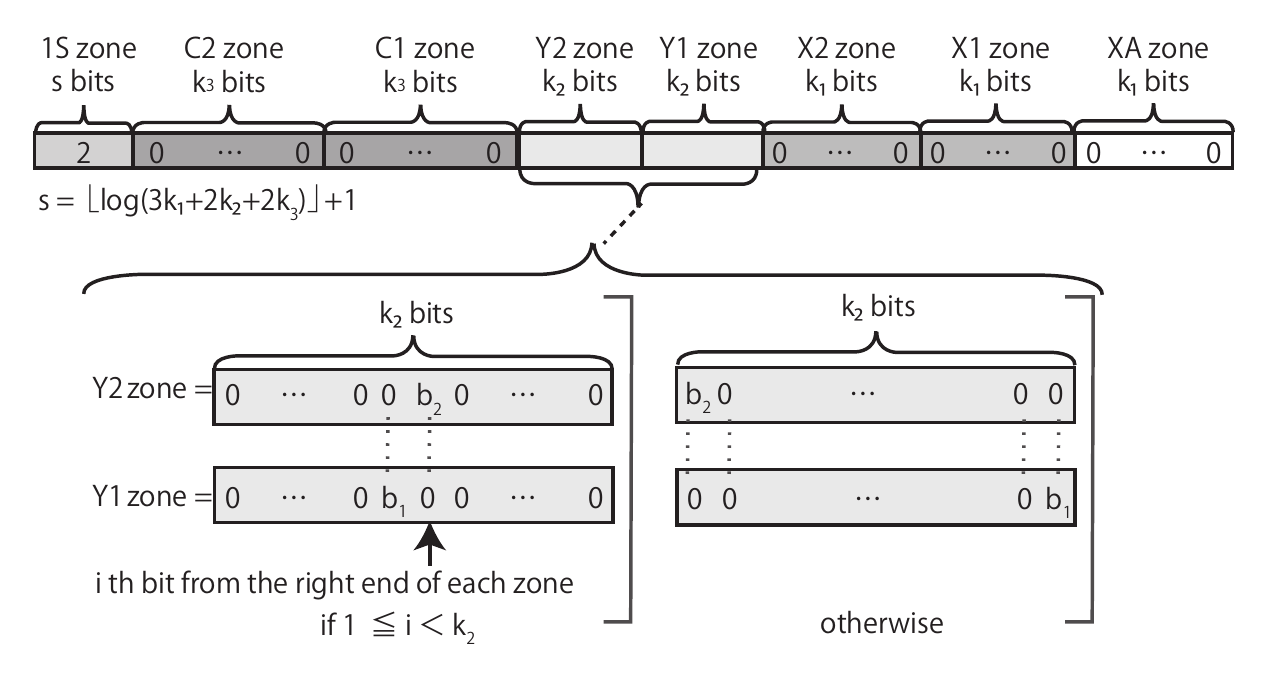}
\end{center}
\caption{Integer $d(y_i,b_2b_1)$ constructed in the proof for Theorem~\ref{thm:3}.}
\label{fig:constructed_integer_for_an adjustment_on_a_variable_in_Y}
\end{figure}

\begin{figure}[htbp]
\begin{center}
\includegraphics{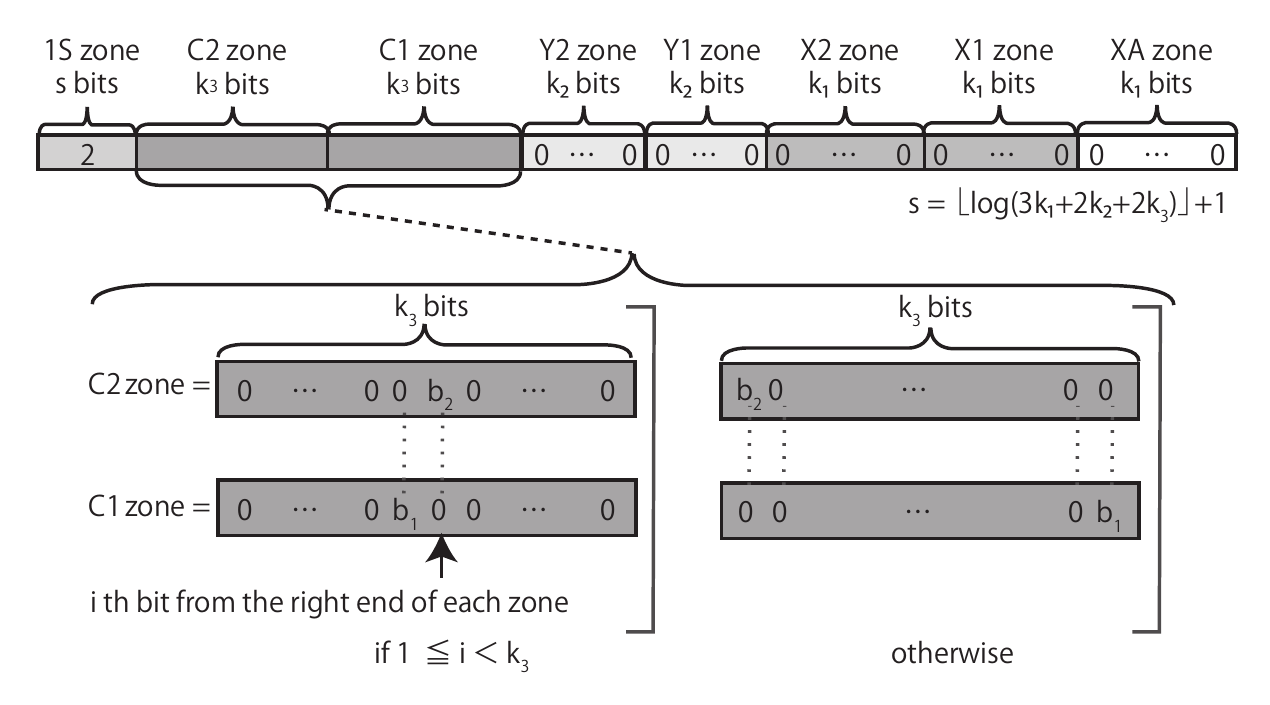}
\end{center}
\caption{The integer $d(C_i,b_2b_1)$ constructed in the proof for Theorem~\ref{thm:3}.}
\label{fig:constructed_integer_for_an adjustment_on_a_clause_in_C}
\end{figure}

For $\varphi$,
we will define a function $s_0$ such that
\begin{eqnarray*}
|\{\sigma_x \in \{0,1\}^{k_1}
\colon (\forall \sigma_y \in \{0,1\}^{k_2})[(2,\varphi,\sigma_x \sigma_y) \in \Phi_{1/3}]\}|
\end{eqnarray*}
is equal to $|N(s_0(\varphi)) \cap [\lambda,\infty]|$,
where $\Phi_{1/3}$ is the set defined in Section~\ref{subsec:computational_problems}.
Then, we will define dummy integers described in Section~\ref{subsec:observation_TGAP}.
For every $x_i \in X_\varphi$, $b_0,b_1,b_2 \in \{0,1\}$,
we define an integer $\DUMMY(x_i, b_2 b_1 b_0)$ as 
\begin{eqnarray*}
3 \DUMMY_0 + 2^{2k_1+i-1}b_2 + 2^{k_1+(i \MOD k_1)}b_1 + 2^{i-1}b_0.
\end{eqnarray*}
Figure~\ref{fig:constructed_integer_for_an adjustment_on_a_variable_in_X} 
illustrates the binary representation of $\DUMMY(x_i, b_2 b_1 b_0)$.
Similarly, for every $y_i \in Y_\varphi$, $b_1,b_2 \in \{0,1\}$,
we define an integer $\DUMMY(y_i, b_2 b_1)$ as 
\begin{eqnarray*}
2 \DUMMY_0 + 2^{3k_1+k_2+i-1}b_2 + 2^{3k_1+(i \MOD k_2)}b_1.
\end{eqnarray*}
Figure~\ref{fig:constructed_integer_for_an adjustment_on_a_variable_in_Y} illustrates
the binary representation of $\DUMMY(y_i, b_2 b_1)$.
For every clause $C_i$ with $1 \leq i \leq k_3$ and $b_1,b_2 \in \{0,1\}$,
we define an integer $\DUMMY(C_i, b_2 b_1)$ as 
\begin{eqnarray*}
2 \DUMMY_0 + 2^{3k_1 + 2k_2 + k_3 + i -1}b_2 + 2^{3k_1+2k_2+(i \MOD k_3)}b_1,
\end{eqnarray*}
Figure~\ref{fig:constructed_integer_for_an adjustment_on_a_clause_in_C} illustrates
the binary representation of $\DUMMY(C_i, b_2 b_1)$.
For every $x \in X_\varphi$,
we denote the set 
$\{\DUMMY(x, \beta) \colon \beta \in \{0,1\}^3 \backslash \{000\}\}$
by $D(x)$.
For every $z \in Y_\varphi \cup \mathcal{C}_\varphi$,
we denote the set 
$\{\DUMMY(z, \beta) \colon \beta \in \{0,1\}^2 \backslash \{00\}\}$
by $D(z)$.

For every $x \in X_\varphi$ and $\beta \in \{0,1\}^3$,
we define $d_+(x,\beta)$ as $d(x,\beta) + \DUMMY_0$; i.e.,
$d_+(x,\beta)$ is the integer
obtained from $d(x,\beta)$ by incrementing its $\OSPT$ zone.
We denote the set 
$\bigcup_{z \in X_\varphi \cup Y_\varphi \cup \mathcal{C}_\varphi} D(z)$ by $D$.
Moreover, we denote the set 
\begin{eqnarray*}
\left(D \backslash \{d(x_{k_1},110),d(x_{k_1},111)\}\right) \cup  \{d_+(x_{k_1},110),d_+(x_{k_1},111)\}
\end{eqnarray*}
by $D_+$.
We define $s_0(\varphi)$ as the set 
\begin{eqnarray*}
H \cup D_+ \cup \{\DUMMY_0\}.
\end{eqnarray*}

By \ref{alg:2}, 
the reduction can compute $(H \cup D_+ \cup \{\DUMMY_0\}, \lambda)$
in polynomial time.

\begin{algorithm}
\caption{Reduction from $\NUMD{\PI{1}\OT\SAT}$ to $\TGAP$}
\label{alg:2}
\begin{algorithmic}
 \STATE {\bfseries Input.} A $3$-CNF formula 
 $\varphi = (C_1 \land \cdots \land C_{k_3})$ over $X \cup Y$, 
 where $|X_\varphi| = k_1$ and $|Y_\varphi| = k_2$.
 \STATE {\bfseries Output.} $(H \cup D_+ \cup \{\DUMMY_0\}, \lambda)$.
  \STATE Execute the same process in \ref{alg:1}
  (After executing this process, we obtain $\DUMMY_0$, $\lambda$, and $H$)
  \FOR{each $l \in \{k_1,k_2,k_3\}$}
   \IF{$l = k_1$}
    \STATE $q \leftarrow k_1$
   \ELSIF{$l = k_2$}
    \STATE $q \leftarrow 3k_1$    
   \ELSE
    \STATE $q \leftarrow 3k_1+2k_2$    
   \ENDIF
   \FOR{each $i = 1$ to $l$}
    \FOR{each $b_2b_1 \in \{0,1\}^q$ such that $b_2b_1 \neq 00$ if $l \neq k_1$} 
     \STATE $v \leftarrow 0$
     \IF{$b_2=1$}
      \STATE $v \leftarrow v + 2^{q + l + i - 1}$
     \ENDIF
     \IF{$b_1=1$}
      \STATE $v \leftarrow v + 2^{q + (i \MOD l)}$
     \ENDIF
     \IF{$l = k_1$} 
      \FOR{each $b_0 \in \{0,1\}$}
       \IF{$b_2b_1b_0 \neq 000$}
	\IF{$i = k_1$ and $b_2b_1 = 11$}
	 \STATE $D_+ \leftarrow D_+ \cup \{v + 4 \DUMMY_0 + 2^{i-1} b_0\}$
	\ELSE
	 \STATE $D_+ \leftarrow D_+ \cup \{v + 3 \DUMMY_0 + 2^{i-1} b_0\}$
	\ENDIF
       \ENDIF
      \ENDFOR     
     \ELSE
      \STATE $D_+ \leftarrow D_+ \cup \{v + 2 \DUMMY_0\}$
     \ENDIF
    \ENDFOR 
   \ENDFOR 
  \ENDFOR 
  \RETURN $(H \cup D_+ \cup \{\DUMMY_0\}, \lambda)$
\end{algorithmic}
\end{algorithm} 

Then, we show the validity of the reduction.
In the rest of the proof for Theorem~\ref{thm:3}, 
let $\alpha_1,\cdots,\alpha_\iota$ be integers in $s_0(\varphi)$ 
and let $\alpha$ be the sum of $\alpha_1,\cdots,\alpha_\iota$.
For every $1 \leq \kappa \leq \iota$,
let $\xi_\kappa$ be $\sum_{i = 1}^ \kappa \alpha_i$.
That is, $\xi_1 = \alpha_1$ and $\xi_\iota = \alpha$.
For any nonnegative integer $e$,
we say that $e$ is {\itshape consistent} with $\varphi$
if $e$ satisfies the same condition as in the proof for Theorem~\ref{thm:2}.
Note that this notion is well-defined for any nonnegative integer.
Moreover, by definition, if $\alpha$ is in $[\lambda,\varphi]$,
then $\alpha$ is consistent with $\varphi$.
We can prove the following claim by the similar discussion
as in the proof for Claim~\ref{cla:4}.
\begin{claim}
\label{cla:7}
 $\xi_i$ is consistent with $\varphi$
 for every $1 \leq i \leq \iota$
 if and only if
 no carry occurs at 
 the $\XAPT$, $\XOPT$, $\XTPT$, $\YOPT$, $\YTPT$, $\COPT$, and $\CTPT$ zones
 in the addition $\xi_{i-1} + \alpha_i$
 for every $2 \leq i \leq \iota$.
\end{claim}
\begin{claim}
\label{cla:11}
If $\alpha \in [\lambda,\mu]$ and there is no $S \subseteq H$
such that $\sum_{\tau \in S} \tau = \alpha$,
then $\alpha \in N(s_0(\varphi))$.
\end{claim}
\noindent{\itshape Proof of Claim~\ref{cla:11}.}
Let $\alpha$ be in $[\lambda,\mu]$.
Let $S_1$ be a subset of $H$ and $S_2$ be a subset of $D_+$
such that $\sum_{\tau \in S_1 \cup S_2} \tau = \alpha$.
By assumption, if $S_2 = \emptyset$, then
$S_1 \subseteq H$
and $\sum_{\tau \in S_1} \tau = \alpha$,
which is a contradiction.
Thus, $S_2$ is nonempty.
Assume that $S_1 \neq \emptyset$ and $\alpha$ is in $[\lambda,\mu]$.
Then, by the definitions of integers in $H$ or $D_+$,
at least one carry occurs in one of the $\XOPT,\XTPT,\YOPT,\YTPT,\COPT,\CTPT$ zones.
This contradicts to Claim~\ref{cla:7}.
\QED (Claim~\ref{cla:11})

\begin{claim}
\label{cla:13}
If $\alpha \geq \mu+1$,
then we can find a set $S \subseteq D_+$
such that $\sum_{\tau \in S} \tau = \alpha$.
\end{claim}
\noindent{\itshape Proof of Claim~\ref{cla:13}.}
By the definitions of integers in $D_+$,
we can represent every integer in $[\mu+1,\mu+\DUMMY_0]$
as a nonnegative integer combination of $D_+$.
Thus, the claim holds. 
\QED (Claim~\ref{cla:13})

\begin{claim}
\label{cla:3}
All integers in $s_0(\varphi)$ are coprime.
\end{claim}
\noindent{\itshape Proof of Claim~\ref{cla:3}.}
By definition, $D$ contains two successive integers;
e.g., $d(x_1,010)$ and $d(x_1,011)$.
By Euclidean algorithm~(\cite{Graham:1994:CMF:562056}, Chapter 4),
two successive integers are coprime.
Thus, all integers in $s_0(\varphi)$ are coprime.
\QED (Claim~\ref{cla:3})

Moreover, we can check that Claim~\ref{cla:2} holds
even if $\alpha$ is defined as an integer
represented as a nonnegative integer combination
of $H \cup D_+ \cup \{\DUMMY_0\}$.
The proof for Theorem~\ref{thm:3} is complete.
\QED (Theorem~\ref{thm:3})

\section{Complexity of $\NUM\GAP$}
\label{sec:proof_relax_subtractive_reducibility}

In this section, we prove Theorem~\ref{thm:4}
by extending the reduction in the proof for Theorem~\ref{thm:3} 
to a reduction from $\NUMD{\PI{1}\OT\SAT}$ to $\NUM\GAP$.
Theorem~\ref{thm:4} is a main theorem in this paper.

\begin{theorem}
\label{thm:4}
$\NUM\GAP$ is $\NUM\NP$-complete under relaxed subtractive reductions.
\end{theorem}

\subsection{Closure property of  $\NUM\DOT\PIP{k}$ under relaxed subtractive reductions}
\label{subsec:proof_closure_relaxed_subtractive_reduction}

In Section~\ref{subsec:2_reducibilities},
we introduced a new type of polynomial-time reduction that we call a relaxed subtractive reduction.
In this subsection, 
we prove that $\NUM\DOT\PIP{k}$ is closed under relaxed subtractive reductions.
The proof takes the same approach as the proofs for
the closure property of $\NUM\DOT\PIP{k}$ under
subtractive reductions in \cite{Durand2005496}
and complementive reductions in \cite{BaulandBohlerCreignouReithSchnoorVollmer2010TOCS}.

\begin{theorem}
\label{thm:1}
For every $k \geq 1$,
the class $\NUM \cdot \PIP{k}$ is closed under
relaxed subtractive reductions.
\end{theorem}
\noindent{\itshape Proof of Theorem~\ref{thm:1}.} 
Let $\NUM A$ and $\NUM B$ be counting problems.
Suppose that $\NUM B \in \NUMD{\PIP{k}}$.
Let a pair $(t_0,t_1)$ of polynomial-time computable functions
be a relaxed subtractive reduction from $\NUM A$ to $\NUM B$
and $F \in \CLASSP$ be a decision problem such that $\NUM F \in \FP$,
which satisfy the following conditions.
For every $w \in \{0,1\}^\ast$,
\begin{enumerate}
\item $W_F(w) \subseteq W_B(t_0(w))$
\item $W_B(t_1(w)) \subseteq W_B(t_0(w))$
\item $W_F(w) \cap W_B(t_1(w)) = \emptyset$
\item $|W_A(w)| = |W_B(t_0(w))| - |W_B(t_1(w))| - |W_F(w)|$.
\end{enumerate}
These $4$ conditions are from the definition 
of a strong relaxed subtractive reduction.
For proving the theorem,
it suffices to show that
\begin{enumerate}[(S1)]
\item there is a decision problem $A^\prime$ such that $R_{A^\prime} \in \CLASSP^{\SIGMAP{k}}$,
\item $|W_{A^\prime}(w)| = |W_A(w)|$ for every $w \in \{0,1\}^\ast$.
\end{enumerate}
By the equality $\NUMD{\PIP{k}} = \NUMD{\CLASSP^{\SIGMAP{k}}}$~\cite{TodaPhD1991English},
the statement (S1) implies that $\NUM A^\prime \in \NUMD{ \PIP{k}}$.
The statements (S1) and (S2) imply that
$\NUM A \in \NUMD{ \PIP{k}}$.

Let $N^{L_B}$ be a deterministic oracle machine
that recognizes $B$, where
$L_B$ is an oracle in $\SIGMAP{k}$.
By the equality $\NUMD{\PIP{k}} = \NUMD{\CLASSP^{\SIGMAP{k}}}$,
$R_B$ is in $\CLASSP^{\SIGMAP{k}}$.
Let $R_{A^\prime}$
consist of all pairs $(w, w_0\DELIM w_1 \DELIM v)$
such that $w_0 = t_0(w)$, $w_1 = t_1(w)$, 
and $v \in (W_B(w_0) \backslash W_B(w_1)) \backslash W_F(w)$,
where $\DELIM $ is a new symbol.
We show that $w_0 \DELIM w_1 \DELIM v$ is a witness of $w$
by using a deterministic oracle machine $M^{L_B}$.
$M^{L_B}$ simulates $N^{L_B}$ as a subroutine
with overhead bounded by a polynomial in the size of $w$.

Let $\gamma$ be $w_0\DELIM w_1\DELIM v$. 
We can extract $w_0$ from $\gamma$ in polynomial time.
Moreover, we can check whether $w_0$ is obtained from $w$ in polynomial time,
since $t_0$ is a polynomial-time computable function.
We can do the same process for $w_1$ in $\gamma$.
After doing these processes,
we extract $(w_0, v)$ from $\gamma$.
We can do this process in polynomial time.
Then, we check that $v$ is in $W_B(w_0)$.
By using $N^{L_B}$ as a subroutine,
we can check that $v$ is in $W_B(w_0)$ in polynomial time.
Similarly, we can execute the following steps in polynomial time.
We extract $(w_1, v)$ from $\gamma$ 
and check that $v$ are not in $W_B(w_1)$.
Finally, we check $(w, v)$ to be not in $R_F$.
We can execute this step in polynomial time
since we assume $\NUM F$ to be in $\FP$.
Thus, $\NUM A^\prime$ is in $\NUM\DOT\PIP{k}$.
By definition, $|W_{A^\prime}(w)|$ is equal to
$|(W_B(t_0(w)) \backslash W_B(t_1(w))) \backslash W_F(w)|$.

By assumption,
$W_{A}(w)$ is $W_B(t_0(w)) \backslash W_B(t_1(w)) \backslash W_F(w)$.
Thus, $|W_{A^\prime}(w)|$ is equal to $|W_A(w)|$.
This implies that $\NUM A^\prime = \NUM A$.
Consequently, $\NUM A$ is in $\NUMD{ \PIP{k}}$.
The proof is complete.
\QED

\subsection{Ideas and examples of the reduction}
\label{subsec:ideas_and_examples}

In this subsection, we describe main ideas
of our reduction from $\NUMD{\PI{1}\OT\SAT}$ to $\NUM\GAP$.
In Section~\ref{sec:proof_bounded_gaps_parsimonious_reducibility},
we proved that 
$\NUMD{\PI{1}\OT\SAT}$ is parsimonious reducible to $\NUM\TGAP$.
This implies that $\NUMD{\PI{1}\OT\SAT}$ is
also relaxed subtractive reducible to $\NUM\TGAP$.
By extending the reduction in the proof for Theorem~\ref{thm:3},
we will construct a strong relaxed subtractive reduction
from $\NUMD{\PI{1}\OT\SAT}$ to $\NUM\GAP$.

Let $\varphi, [\lambda, \mu], H, h$, and $s_0$ 
be as in the proof for Theorem~\ref{thm:3}. 
Our goal is obtaining a function $s_1$ and a problem $F \in \CLASSP$,
which satisfy the following conditions.
\begin{enumerate}[(C1)]
 \item Every integer in $N(s_0(\varphi)) \cap [\lambda, \mu]$
 is in $N(s_0(\varphi)) \backslash N(s_1(\varphi))$.
 \item $F$ coincides $(N(s_0(\varphi)) \backslash N(s_1(\varphi))) \cap [1,\lambda-1]$.
\end{enumerate}
As a result,
we want to construct a (strong) relaxed subtractive reduction
$(s_0,s_1)$ such that
\begin{eqnarray*}
|N(s_0(\varphi)) \cap [\lambda, \infty]|  
= |N(s_0(\varphi))| - |N(s_1(\varphi))| - |W_F(\varphi)|.
\end{eqnarray*}

\begin{figure}[htbp]
\begin{center}
\includegraphics{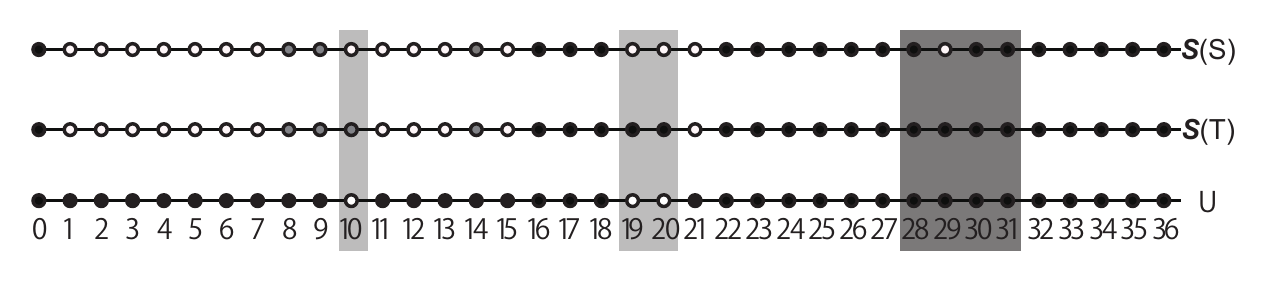}
\end{center}
\caption{Gaps and nongaps of numerical semigroups generated by $S,T,U$ and a set $F$,
where $S = \{8,9,14\}$, $T = \{8,9,10,14\}$, $U = \{2,7\}$, $F=\{10,19,20\}$}
\label{fig:3}
\end{figure}

Let us informally describe 
the relaxed subtractive reduction $(s_0,s_1)$
by using examples.
For observing essential properties of numerical semigroups for our reduction,
we first consider a simple example such that 
a pair of a set of positive integers and an interval are given
but may not be constructed from a $3$-CNF formula.
Let $(S, [28,31])$ be given, where $S= \{8,9,14\}$.
We simulate every integer in $N(S) \cap [1,27]$ 
as an integer in $N(T) \cup U$ as in Figure~\ref{fig:3},
where $T=\{8,9,10,14\}$ and $U=\{10,19,20\}$.
$U$ coincides with $N(S) \backslash N(T)$.
In Figure~\ref{fig:3},
every circle means an integer.
The $1$st two lines correspond to the numerical semigroups generated by $S$ and $T$, respectively.
On these two lines,
every black circle means a nongap of $\mathcal{S}(S)$ or $\mathcal{S}(T)$.
Every gray circle means an element of $S$ or $T$.
Every white circle means a gap of $\mathcal{S}(S)$ or $\mathcal{S}(T)$.
The $3$rd line corresponds to $U$.
On this line, every white circle means an element of $U$.

We can specify all integers of $N(S) \cap [1,27]$ 
by its two partitions.
This partitions are defined by depending on whether an integer is in $N(T)$.
In Figure~\ref{fig:3},
$[1,7] \cup [11,13] \cup \{15,21\}$ 
coincides with $N(S) \cap [1,27] \cap N(T)$. 
$U$ coincides with $(N(S) \cap [1,27]) \backslash N(T)$.

Let us next consider an instance of $\NUM\GAP$,
which is defined for an instance of $\NUMD{\PI{1}\OT\SAT}$.
Let $\varphi_2$ be the $3$-CNF formula 
as in \ref{subsubsec:instance_reduction_TGAP}.
Let $s_0(\varphi_2)$ be a set consisting of $\DUMMY_0$ 
and all integers in 
Tables \ref{tab:A_for_varphi_two}, \ref{tab:D_for_varphi_two} 
or \ref{tab:D_tilde_from_varphi_two}.
Let $s_1(\varphi_2)$ be a set 
consisting of $\DUMMY_0$ and all integers in
Tables \ref{tab:A_for_varphi_two}, \ref{tab:D_for_varphi_two}, %
\ref{tab:D_plus_xk1_for_varphi_two}, or \ref{tab:D_tilde_from_varphi_two}.
Then, we specify
all integers in $N(s_0(\varphi_2)) \cap [\lambda_{\varphi_2}, \infty]$
as follows.
Since $s_0(\varphi_2)$ is a subset of $s_1(\varphi_2)$,
$N(s_0(\varphi_2))$ contains $N(s_1(\varphi_2))$ as a subset; i.e.,
any integer that cannot be represented as a nonnegative integer combination of $s_0(\varphi_2)$
cannot be done as of $s_1(\varphi_2)$.
Thus, the task that is left
is a specification of $(N(s_0(\varphi_2)) \backslash N(s_1(\varphi_2))) \cap [1,\lambda-1]$.
Let $F$ be $(N(s_0(\varphi_2)) \backslash N(s_1(\varphi_2))) \cap [1,\lambda-1]$.
Moreover, we can observe that
$N(s_1(\varphi_2))$ does not include
any integer of $N(s_0(\varphi)) \cap [\lambda_{\varphi_2}, \infty]$
by checking each bit in the binary representation of 
each integer in $s_1(\varphi_2)$ and the integer $\lambda_{\varphi_2}$.

\begin{table}[htbp]
\caption{Integers constructed from $\varphi_2$, which are introduced for the reduction to $\GAP$.}
\label{tab:D_tilde_from_varphi_two}
\begin{center}
\begin{tabular}{cc|cccccccc}
\hline
Variables & \multirow{2}{*}{Values}
& \multicolumn{8}{c}{Constructed integers} \\
\cline{3-10}
or clauses & & $\OSPT$ & $\CTPT$ & $\COPT$ 
& $\YTPT$ & $\YOPT$ & $\XTPT$ & $\XOPT$ & $\XAPT$
\\
\hline
\multirow{2}{*}{$x_2$} & $110$ & $3$ & $00$ & $00$ & $00$ & $00$ & $10$ & $01$ & $00$  \\
 & $111$ & $3$ & $00$ & $00$ & $00$ & $00$ & $10$ & $01$ & $10$  \\
\hline
\end{tabular}
\end{center}
\end{table}

\subsection{Formal discussion for Theorem~\ref{thm:4}}
\label{subsec:main_proof}

\noindent
{\itshape Proof of Theorem~\ref{thm:4}.} 
By definition $\NUM\GAP$ is a special case of $\NUM\TGAP$.
Thus, by Theorem~\ref{thm:3}, $\NUM\GAP$ is in $\NUM\NP$.
All that is left is to prove $\NUM\GAP$ to be $\NUM\NP$-hard 
under relaxed subtractive reductions.

\begin{figure}[htbp]
\begin{center}
\includegraphics{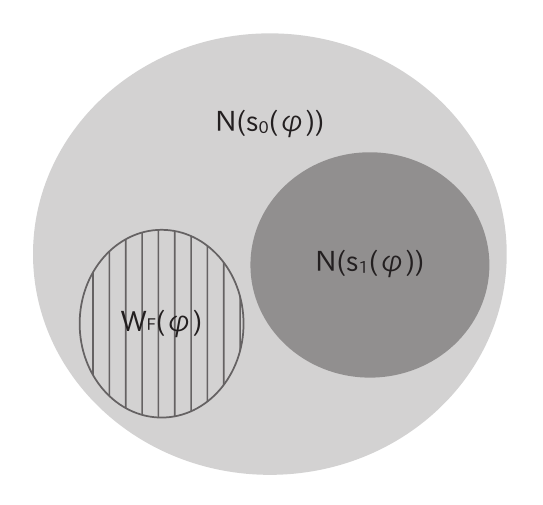}  
\end{center}
\caption{Euler diagram for the strong relaxed subtractive reduction that we construct 
in the proof for Theorem~\ref{thm:4}.}
\label{fig:euler_diagram_strong_relaxed_subtractive_reduction}
\end{figure}

Let $\varphi$ be given as in the proof for Theorem~\ref{thm:3}. 
Let $d$, $[\lambda, \mu]$, $H$, $h$, and $s_0$ 
be defined as in the proof for Theorem~\ref{thm:3}. 
We will define a function $s_1$ and find a decision problem $F$ 
such that 
\begin{enumerate}[(C1)]
 \item  $N(s_1(\varphi)) \subseteq N(s_0(\varphi))$,
 \item  $W_F(\varphi) \subseteq N(s_0(\varphi))$,
 \item  $N(s_1(\varphi)) \cap W_F(\varphi) = \emptyset$,
 \item  There is a $\sigma_x \in \{0,1\}^{|X_\varphi|}$ 
such that $(k,\varphi, \sigma_x \sigma_y) \not\in \Phi_{1/3}$
for every $\sigma_y \in \{0,1\}^{|Y_\varphi|}$
if and only if there is an integer in
$\left(N(s_0(\varphi)) \backslash N(s_1(\varphi)\right) \backslash W_F(\varphi)$.
\end{enumerate}
Figure~\ref{fig:euler_diagram_strong_relaxed_subtractive_reduction}
illustrates relationships of $N(s_0(\varphi))$, $N(s_1(\varphi))$, and $W_F(\varphi)$.
Let $\tilde{D}_+$ be the set
$D_+ \cup \{d(x,110), d(x,111)\}$.
$\tilde{D}_+$ coincides with 
$D \cup \{d_+(x,110), d_+(x,111)\}$. 
We define $s_1(\varphi)$ as 
\begin{eqnarray*}
& H \cup \tilde{D}_+ \cup \{\DUMMY_0\}, 
\ \ \ \ \ \text{i.e.,} \ \ \ \ \ 
s_0(\varphi) \cup \{d(x_{k_1},110), d(x_{k_1},111)\}. 
\end{eqnarray*}
We can check that $s_1$ can be constructed in polynomial time
by the same discussion as for $s_0$ in the proof of Theorem~\ref{thm:3}.

Next, we define $F$ as a decision problem.
We define the polynomially balanced binary relation $R_F$ as follows.
A pair $(\varphi, v)$ is in $R_F$ if and only if
$f$ is a $3$-CNF formula
and $v$ is the sum of all positive integers in a set $B_\varphi$. 
$B_\varphi$ satisfies the following conditions.
\begin{itemize}
\item Exactly one of $d_\varphi(x_{k_1},110)$ and $d_\varphi(x_{k_1}, 111)$
is in $B_\varphi$.
\item For every $x \in X_\varphi \backslash \{x_{k_1}\}$, 
at most one element in $\{d_\varphi(x,\beta) \colon \beta \in \{0,1\}^3, \beta \neq 000\}$
is in $B_\varphi$.
\item For every $z \in Y_\varphi \cup \mathcal{C}_\varphi$, 
at most one element in $\{d_\varphi(z,\beta) \colon \beta \in \{0,1\}^2, \beta \neq 00\}$
is in $B_\varphi$.
\end{itemize}
Then, an instance of $F$ is a $3$-CNF $f$ over $X \cup Y$,
the question of $F$ is whether there is an integer $v$ such that $(\varphi,v) \in R_F$.
By definition, $F$ is in $\CLASSP$.
Moreover, $\NUM F(\varphi)$ is $2 \cdot 8^{k_1-1} \cdot 4^{k_2+k_3}$; i.e., $2^{3k_1+2k_2+2k_3-2}$,
which is computable in time polynomial in $k_1,k_2$, and $k_3$.
Thus, the counting problem $\NUM F$ is in $\FP$.

In the rest of the proof for Theorem~\ref{thm:4},
we will show the validity of the above construction.
Let $\alpha_1,\cdots,\alpha_\iota$ be integers in $s_1(\varphi)$
and let $\alpha$ be the sum of $\alpha_1,\cdots,\alpha_\iota$.
For every $1 \leq \kappa \leq \iota$,
let $\xi_i$ be $\sum_{i = 1}^ \kappa \alpha_i$.
For any nonnegative integer $e$,
we say that $e$ is {\itshape consistent} with $\varphi$
if $e$ satisfies the same condition as in the proof for Theorem~\ref{thm:3}.
We can prove the following claim by the similar discussion
in the proof for Theorems Theorem~\ref{thm:2} and Theorem~\ref{thm:3}.
\begin{claim}
\label{cla:8} 
 $\xi_i$ is consistent with $\varphi$
 for every $1 \leq i \leq \iota$
 if and only if
 no carry occurs at the $\XAPT$, $\XOPT$, $\XTPT$, $\YOPT$, $\YTPT$, $\COPT$, and $\CTPT$ parts
 in the addition $\xi_{i-1} + \alpha_i$
 for every $2 \leq i \leq \iota$.
\end{claim}
By Claim~\ref{cla:8}, we can verify Claims \ref{cla:12}, \ref{cla:9}, and \ref{cla:6}.
\begin{claim}
\label{cla:12}
If $\alpha \geq \lambda$,
then we can find a set $S \subseteq \tilde{D}_+ \cup \{\DUMMY_0\}$
such that $\sum_{\tau \in S} \tau = \alpha$.
\end{claim}

\begin{claim}
\label{cla:9}
If $\alpha \in [\lambda,\mu]$ and there is no $S \subseteq H$
such that $\sum_{\tau \in S} \tau = \alpha$,
then $\alpha \in N(s_0(\varphi))$.
\end{claim}

\begin{claim}
\label{cla:6}
Let $\alpha$ be less than $\lambda$.
$\alpha \in W_F(\varphi)$ if and only if 
there is an $K \subseteq s_1(\varphi)$ such that $\sum_{\tau \in K} \tau = \alpha$  
and there is no $L \subseteq s_0(\varphi)$ such that $\sum_{\tau \in L} \tau = \alpha$
\end{claim}

We can observe the validities of the statements of 
Claims \ref{cla:12}, \ref{cla:9}, and \ref{cla:6}
by using examples of Figure~\ref{fig:3}.
Let $S$, $T$, and $U$ be as in Section~\ref{subsec:ideas_and_examples}.
$S,T$ and $U$ correspond to $s_0(\varphi)$, $s_1(\varphi)$, and $W_F(\varphi)$, respectively.
Claim~\ref{cla:12} corresponds to the fact that 
every integer greater than or equal to
the lower endpoint $28$ can be represented 
as a nonnegative integer combination of $T$.
Claim~\ref{cla:9} corresponds to the fact that
$29$ cannot be represented as a nonnegative integer combination of $S$.
Claim~\ref{cla:6} corresponds to the fact that
$10, 19$ and $20$ are in $U$
and can be represented as nonnegative integer combinations of $T$
but cannot of $S$. 
The following claim follows from Claim~\ref{cla:3}.

\begin{claim}
\label{cla:5}
All integers in $s_1(\varphi)$ are coprime.
\end{claim}

By definition,
the pair $(s_0,s_1)$ satisfies 
the conditions of strong relaxed subtractive reductions.
The proof for Theorem~\ref{thm:4} is complete.
\QED (Theorem~\ref{thm:4})

\section{Future work}
\label{sec:future_work_and_concluding_remarks}

In this paper, we proved the $\NUM\NP$-completeness of $\NUM\GAP$ 
under relaxed subtractive reductions as a main result.
A relaxed subtractive reduction is new type of polynomial-time reduction.
Moreover,
we proved the $\NUM\NP$-completenesses of $\NUM\IGAP$ and $\NUM\TGAP$, 
which are variants of $\NUM\GAP$, 
under parsimonious reductions.
This section describes future work.

\subsection{$\NUM\NP$-hardness of  $\NUM\GAP$ under parsimonious reductions}
\label{subsec:future_work_1}

We showed the $\NUM\NP$-completeness of $\NUM\GAP$ 
under relaxed subtractive reductions.
It is natural to ask whether $\NUM\GAP$ is $\NUM\NP$-complete
under parsimonious reductions.
Unfortunately, it appears to be quite difficult to answer this question.

In this paper, we mainly use combinatorial and logical approaches for constructing reductions,
but use little result in number theory.
Thus, it may be possible to prove $\NUM\GAP$ to be $\NUM\NP$-complete
under parsimonious reductions by using some tools developed in number theory.
Cook~\cite{Cook:1971:CTP:800157.805047} said that 
the primality testing problem requires a deep insight on number theory.
Indeed, over thirty years later,
\cite{AgrawalKayalSaxena2004} found a polynomial-time algorithm 
by an approach from number theory.

\subsection{Proof methods for the $\NUM\NP$-hardness of $\NUM\TGAP$}
\label{subsec:discussion_TGAP}

To prove the $\NUM\NP$ hardness of $\NUM\TGAP$,
we reduced $\NUMD{\PI{1}\OT\SAT}$ to $\NUM\TGAP$.
However, a reduction from $\NUM\IGAP$ to $\NUM\TGAP$
is another natural reduction.
Since the latter reduction is a self-reduction,
it appears to be more natural than our reduction.
Nevertheless, we could not adopt this approach by the following obstacle.

A parsimonious reduction from $\NUM\IGAP$ to $\NUM\TGAP$
requires a deep insight into number theoretical aspects of numerical semigroups.
Given an instance $(A^\prime,[\lambda, \mu])$ of $\NUM\IGAP$,
no polynomial-time algorithm is known 
for finding an instance $(A, \kappa)$ of $\NUM\TGAP$
such that 
\begin{enumerate}[(C1)]
 \item $|N(A) \cap [\kappa, \infty]| 
 = |N(A^\prime) \cap [\lambda, \mu]|$ and
 \item the size of $(A, \kappa)$
 is of polynomial order in the size of 
 $(A^\prime,[\lambda, \mu])$.
\end{enumerate}
To the best of the author's knowledge,
little relationship is known
for an instance of $\NUM\TGAP$ and an instance of $\NUM\IGAP$.
Our reduction in Section~\ref{sec:proof_bounded_gaps_parsimonious_reducibility} makes use of
properties originated from an instance of $\NUMD{\PI{1}\OT\SAT}$.

\subsection{Completeness for $\NUMD{\PIP{k}}$ under more general reductions}
\label{subsec:future_work_3}

We introduced relaxed subtractive reductions
as a generalization of subtractive reductions.
Moreover, relaxed subtractive reductions are also
special case of polynomial-time bounded truth-table reductions~\cite{LADNER1975103}.
As a property of this type of reduction,
relaxed subtractive reductions
have nonadaptivity; i.e.,
all oracle queries must be decided before starting the other processes.
Moreover, in every relaxed subtractive reduction, 
its normal processes are also independent of its two oracle queries.
We can consider that 
a $\NUMD{\PIP{k}}$-complete problem exists
under relaxed subtractive reductions
due to this property.
However, we do not know 
whether there is a more general type of nonadaptive reduction under which $\NUMD{\PIP{k}}$ is closed.
This is an interesting subject since
$\NUMD{\PIP{k}}$ is not closed under polynomial-time $1$ Turing reductions~\cite{TODA1992205}.

\end{document}